\documentclass[11pt,a4paper]{article}
\usepackage[margin=1.3in]{geometry}
\newcommand{\Rone}[1]{#1}
\newcommand{\Rtwo}[1]{#1}
\newcommand{\Rzero}[1]{#1}
\newcommand{\appendixref}[1]{\autoref{#1}}

\usepackage[english]{babel}
\usepackage{amssymb}
\usepackage[dvipsnames]{xcolor}
\usepackage{graphicx}
\usepackage{graphbox} % allows includegraphics[align=c]

\usepackage{mathtools}
\usepackage[font=normalsize]{caption}
\usepackage[font=normalsize,margin=2mm]{subfig}
\usepackage{float}
\usepackage{hyperref}
\hypersetup{colorlinks=true,
		linkcolor=blue,
		filecolor=blue,
		urlcolor=blue,
		citecolor=blue
	}
\usepackage{xfrac}
\usepackage{enumitem}
\usepackage{gensymb}
\usepackage{tikz}
\usetikzlibrary{patterns}		
\usepackage{listings} % code envirorment
\input{listingsEnvirorment.sty}

\usepackage[authoryear,round]{natbib}

\newcommand{\mr}{\mathrm}
\newcommand{\mc}{\mathcal}
\let\SSS\S
\renewcommand{\S}{^\mr{s}}
\newcommand{\ii}{\mr{i}\,}
\newcommand{\ee}{\mr{e}}
\let\underscore\_
\renewcommand{\_}[1]{_\mr{#1}}
\let\slashHat\^
\renewcommand{\^}[1]{^\mr{#1}}
\let\Re\relax
\let\Im\relax
\DeclareMathOperator\Re{Re}
\DeclareMathOperator\Im{Im}
\DeclareMathOperator\sech{sech}
\DeclareMathOperator\csch{csch}
\newcommand{\h}{\hat}
\newcommand{\br}[3]{\left#1#2\right#3}
\newcommand{\rbr}[1]{\left(#1\right)}
\newcommand{\sbr}[1]{\left[#1\right]}
\newcommand{\cbr}[1]{\left\{#1\right\}}
\newcommand{\dd}{\mr d}
\newcommand{\ddfrac}[2]{\frac{\dd #1}{\dd #2}}

\newcommand{\mean}[1]{\br\langle{#1}\rangle}

\newcommand{\zzvar}{\bar}
\newcommand{\zzzvar}[1]{\bar{\bar{#1}}}

\newcommand{\zzmap}{\zzzvar f}
\newcommand{\zmap}{\zzvar f}

\newcommand{\zDomain}{\mc Z}
\newcommand{\zBoundary}{\partial\zDomain}
\newcommand{\zBed}{\zBoundary\^b}
\newcommand{\zWall}{\zBoundary\^w}
\newcommand{\zSurface}{\zBoundary\^s}

\newcommand{\zzDomain}{\zzvar{\zDomain}}
\newcommand{\zzBoundary}{\partial\zzDomain}
\newcommand{\zzWall}{\zzBoundary\^w}
\newcommand{\zzBed}{\zzBoundary\^b}
\newcommand{\zzSurface}{\zzBoundary\^s}

\newcommand{\zzzDomain}{\zzzvar{\zDomain}}
\newcommand{\zzzBoundary}{\partial\zzzDomain}
\newcommand{\zzzSurface}{\zzzBoundary\^s}

\newcommand{\x}{x}
\newcommand{\y}{y}
\newcommand{\z}{z}
\newcommand{\xx}{{\zzvar x}}
\newcommand{\yy}{{\zzvar y}}
\newcommand{\zz}{{\zzvar z}}
\newcommand{\xxx}{{\zzzvar x}}
\newcommand{\yyy}{{\zzzvar y}}
\newcommand{\zzz}{{\zzzvar z}}

\renewcommand{\h}{h}
\renewcommand{\L}{L}
\newcommand{\w}{w}
\newcommand{\W}{W}
\newcommand{\etaz}{\y\S}
\newcommand{\hh}{{\zzvar \h}}
\newcommand{\LL}{{\zzvar \L}}
\newcommand{\ww}{\zzvar \w}
\newcommand{\WW}{\zzvar \W}
\newcommand{\pphi}{\zzvar \phi}
\newcommand{\UU}{\zzvar{\mc U}}%
\newcommand{\etazz}{\yy\S}
\newcommand{\hhh}{{\zzzvar \h}}

\newcommand{\www}{\zzzvar \w}
\newcommand{\ppphi}{\zzzvar \phi}
\newcommand{\UUU}{\zzzvar{\mc{U}}}
\newcommand{\etazzOfxxx}{\zzzvar\eta}
\newcommand{\fun}{\mu}
\newcommand{\fffun}{\zzzvar\fun}

\newcommand{\mus}{\zzzvar \mu\^s}

\newcommand{\mub}{\zzzvar \mu\^b}

\newcommand{\FF}{\mc F}
\newcommand{\FFInv}{\FF^{-1}}
\newcommand{\kk}{k} % wave number
\newcommand{\MM}{M} % Fourier sum range
\newcommand{\mapC}{\mc C}
\newcommand{\mapS}{\mc S}
\newcommand{\CC}[2]{\sbr{{\mapC_{#1}\!*#2}}}
\renewcommand{\SS}[2]{\sbr{\mapS_{#1}\!*#2}}
\newcommand{\kd}{k\_d}
\newcommand{\kMax}{k\_{max}}
\newcommand{\rDamping}{r}
\newcommand{\N}{n}

\newcommand{\NSC}{J} % N_\angle
\newcommand{\xxjSC}{\zzvar\xi_j}
\newcommand{\thSC}{\theta}

\DeclareMathOperator{\arccosh}{arccosh}

	\title{A precise conformally mapped method for water waves in complex transient environments}
\author{Andreas H. Akselsen\\\textit{\small SINTEF Ocean, Department of Ship and Ocean Structures}\thanks{Paul Fjermstads vei 59, 7052 Trondheim, Norway.}}
\date{February 2025}

\begin{document}

\maketitle

\begin{abstract}
	A two-dimensional water wave model based on conformal mapping is presented.
	The model is exact in the sense that it does not rely on truncated series expansions, nor suffer any numerical diffusion. 
	 Additionally, it is computationally highly  efficient as it numerically evaluates only the surface line while using a fixed number of FFT operations per time step. 
	A double layered mapping  enforces prescribed outer boundaries without iteration.
	The model also supports transient boundaries, including walls.
	Mapping models are presented that support smooth bathymetries and angled overhanging geometries.
	An exact piston-type wavemaker model demonstrates the method's potential as a numerical wave tank.
	The model is tested and validated through a number of examples covering shallow water waves, wavemaker generation, rising bathymetry shelves, and  wave reflection from slanting structures.
	\Rzero{A paddle-type wavemaker model, developed from the present theory, will be detailed in a forthcoming paper.}
\end{abstract}

\section{Introduction}
\label{sec:introduction}
Numerical modelling is taking an ever-larger role in theoretical and applied water wave studies. 
Although often seen as substitute for physical experiments, numerical models frequently feature in a complimentary capacity,
providing validation and a supplementary information. 
In fact, \textit{digital twins} of this kind are not limited to just mimicking experiments  but can be incorporated into them;
time-reversed simulation is a promising  new wavemaker control strategy for reproducing extreme events in a controlled environment \citep{houtani2018_experimentFromHOS,ducrozet2020_timeReversalExperimentalRouge}. 

In numerical wave tanks, both the free surface and transient boundaries are to be modelled.
Commonly applied methods include the Boundary Element Methods (BEM) \citep{khait2019_BEMforWavemakerControl}, Finite Difference and Finite Element Methods (FDM/FEM) using dynamic  sigma-gridding  \citep{wu1995_FEM,wang2022_REEFFNPF},
and Higher Order Spectral methods (HOS) \citep{bonnefoy2006A_BM}. 
The latter provides superior efficiency at approximate accuracy. %and best resembles the method described here. 

Complex boundary shapes manifest in both coastal environments  and basins alike, be it in the form of naturally varying bathymetries or structural components and machinery.
BEM's and sigma-gridded methods support most domain shapes but at lower computationally efficient than  HOS methods.
HOS methods can in turn be made to approximately represent bathymetry variations at some expense to efficiency \citep{gouni2016_HOSbathymetry,gouni2017_HOSbathymetry3D}. 
\\

\Rone{
	Historically, conformal mapping has played a crucial role in the study of hydrodynamics \citep{lamb1932hydrodynamics,milneThomson1962theoreticalHydrodynamics}, numerical computing power being limited. 
	In the context of free-surface flows, it has often been used to characterise surface undulations caused by flow \citep[e.g.,][]{forbes1982_flowOverCylinder,dias1989_flowOverTriangle}.
	More recently, a highly efficient and precise algorithm for describing solitary waves has been developed using conformal mapping 	\citep{clamond2013fast,dutykh2014efficient,clamond2015plethora} and later extended to model Stokes waves at arbitrary water depths \citep{dutykh2016efficient,clamond2018accurate}. 
	This open-source algorithm surpasses traditional perturbation theory solutions for Stokes and cnoidal waves in both accuracy and range of validity and has been extensively used in this work to set up initial simulation conditions.
	}%

\Rtwo{The model presented here builds on the flat-bedded conformal mapping approach of \citet{chalikov1996},
	 further developed in \citet{chalikov2005modeling,chalikov2016Book,chalikov2020} and  \citet{zakharov2002_conformalHOS}.
}
Depth variation was incorporated in a similar model by \citet{viotti2014conformal}  that uses a  fixed-point iteration scheme to simultaneously resolve the unknown surface and the prescribed bathymetry in a single mapping. 
However, this iterative approach significantly reduced efficiency.
\Rone{A key novelty of the  present model is the use of a double-layered mapping
	as illustrated in \autoref{fig:fSketch}.
	By separating prescribed boundary coordinates from the dynamic free-surface coordinates,this approach eliminates the need for iteration while maintaining exactness.}
The resulting method is found to resemble the one by \citet{ruban2004,ruban2005}
who  effectively also  adopts a  double mapping.

Domain shape and motions are prescribed by design of the first mapping layer.
We outline two families of prescribed mappings, one based on Fourier decomposition, suitable for smooth single-valued bathymetries,  and another based on the Schwarz--Christoffel theorem, which accommodates sharp angles.
The latter particularly useful for modelling structural features and supports overhanging geometries, provided they remain topologically connected to the outer boundary. 
Mappings may be transient, allowing for the representation of moving machinery or evolving geophysical features.
\\

Conformal mapping enables concise evaluation of time-dependent boundaries, 
seemingly ideal for representing a moving environment with versatility, accuracy \textit{and}  efficiency.
Like the HOS method, it makes use of the Fast Fourier Transformation algorithm (FFT) while representing the interior domain analytically, requiring computation only along the free surface line.
Unlike HOS, however, it retains exact formal accuracy  without relying on truncated series expansions or suffering from numerical diffusion. 
Higher levels of nonlinearity do not necessitate additional model terms, making conformal mapping particularly advantageous in shallow waters where nonlinear effects are strong.

\Rone{
A fundamental limitation of conformal mapping, unlike HOS methods,   is its restriction to two-dimensional wave fields.
While ideal for wave flumes and long-crested waves, two-dimensional modelling cannot fully capture the long-term evolution of directional wave fields  \citep{chalikov2013_3dModel,chalikov2016Book}. 
The difference affects long-term energy transfer across length scales (analogue to turbulence) and  extreme wave statistics \citep{hennig2015_shortCrestedWaveStatistics}.
The wave breaking behaviour is also different in short-crested waves
 \citep{mcallister2024_3DWaveBreaking} and the structural loads they produce.
While three-dimensional conformal mapping is theoretically possible, it is s far more mathematically complex and  rarely pursued. 
Instead, boundary integral methods, which approximate three-dimensional conformal transformations, are more commonly employed.
}
\\

This paper aims to demonstrate that, despite its abstraction, the conformal mapping strategy is highly applicable to practical studies of wave dynamics and numerical wave tank simulations.
 To illustrate its versatility, 
a range of examples are presented, including shallow-water waves, breaking waves, interactions with complex geometries, transient bathymetries, and wave generation by wavemakers.
These are all detailed in \autoref{sec:ex}, following the wave model in \autoref{sec:model} and domain mapping models in \autoref{sec:map}.
Closing remarks are provided in \autoref{sec:summary}.

\section{The double-layered Conformally Mapped Model}
\label{sec:model}

\subsection{Mapping description}
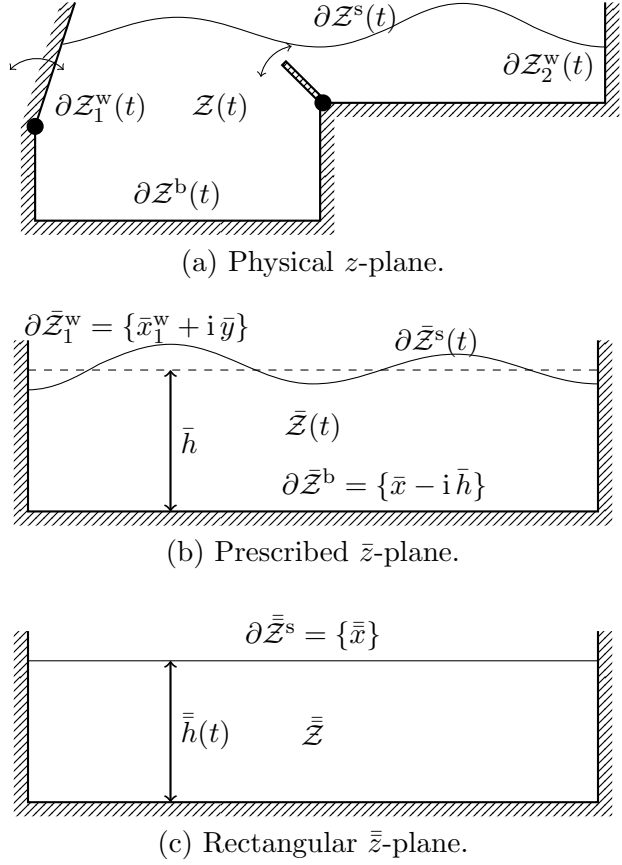
\begin{figure}[H]
	\centering
	\subfloat[Physical $\z$-plane.]{%\documentclass{article}
%\usepackage{tikz}
%\usetikzlibrary{patterns}
%\usepackage{amsmath,bm,amsfonts,amssymb}
%
%
%
%\newcommand{\zDomain}{\mc Z}
%\newcommand{\zzDomain}{\zzvar{\mc Z}}
%\newcommand{\zzzDomain}{\zzzvar{\mc Z}}
%\newcommand{\zBed}{\partial\zDomain\^b}
%\newcommand{\zWall}{\partial\zDomain\^w}
%\newcommand{\zSurface}{\partial\zDomain\^s}
%\newcommand{\zzWall}{\partial\zzDomain\^w}
%\newcommand{\zzBed}{\partial\zzDomain\^b}
%\newcommand{\zzSurface}{\partial\zzDomain\^s}
%\newcommand{\zzzSurface}{\partial\zzzDomain\^s}
%\newcommand{\zBoundary}{\partial\mc Z}
%
%\newcommand{\z}{z}
%\newcommand{\etaz}{{y^{\mathrm s}}}
%\newcommand{\x}{x}
%\renewcommand{\z}{x}
%\newcommand{\w}{w}	
%			
%\begin{document}

%%%%%%%%%%%%%%%%%%%%%%%%%%%%%%%%%%%%%%%%%%

\begin{tikzpicture}[scale=.75]%,font=\large]
	
	% visual
	\def\L{10}
	\def\lam{5} % wavelength
	\def\hd{1/3*\L} % shallow side depth
	\def\hs{1/8*\L} % deep side depth
	\def\LL{1/15*\L} % flap length
	\def\dL{.1}	% flap width
	
	\def\Lh{1/2*\L} % gap width/half length
	\def\wy{.07*\Lh} % wave height
	\def\wx{.25*\lam} % wave quarter length
	\def\tWidth{.05*\Lh} % width of surface texture pattern
	
	% crop image
	\clip (-2*\tWidth,3*\wy) rectangle (\L+\tWidth,-\hd-\tWidth);
	
	% drawin bed
%	\draw[thick]  (0,{1.5*\wy})--(0,-\hd)--(\Lh,-\hd)--(\Lh,-\hs)  --  (\Lh-\LL,-\hs+\LL) --  ({\Lh-(1-\dL)*\LL},{-\hs+(1+\dL)*\LL})  --  (\Lh+2*\dL*\LL,-\hs)   --(2*\Lh,-\hs)--(2*\Lh,{1.5*\wy});
%	\fill[pattern=north east lines, pattern color=black]  (0,-\hd)--(\Lh,-\hd)--(\Lh,-\hs)
%	 -- (\Lh-\LL,-\hs+\LL) -- ({\Lh-(1-\dL)*\LL},{-\hs+(1+\dL)*\LL})  --  (\Lh+2*\dL*\LL,-\hs) 
%	--(2*\Lh,-\hs)--(2*\Lh,1.5*\wy)--(2*\Lh+\tWidth,1.5*\wy)--(2*\Lh+\tWidth,-\hs-\tWidth)--(\Lh+\tWidth,-\hs-\tWidth)--(\Lh+\tWidth,-\hd-\tWidth)--(-\tWidth,-\hd-\tWidth)--(-\tWidth,{1.5*\wy})--(0,{1.5*\wy});
%	\fill[fill=black] (\Lh+.5*\dL,-\hs) circle (1.5*\dL);

	% wall with bend
	\draw[thick] (2*\wy,1.5*\wy)-- (0,-.5*\hd)--(0,-\hd)--(\Lh,-\hd)--(\Lh,-\hs)  --  (\Lh-\LL,-\hs+\LL) --  ({\Lh-(1-\dL)*\LL},{-\hs+(1+\dL)*\LL})  --  (\Lh+2*\dL*\LL,-\hs)   --(2*\Lh,-\hs)--(2*\Lh,{1.5*\wy});
	\fill[pattern=north east lines, pattern color=black]  (0,-\hd)--(\Lh,-\hd)--(\Lh,-\hs)
	-- (\Lh-\LL,-\hs+\LL) -- ({\Lh-(1-\dL)*\LL},{-\hs+(1+\dL)*\LL})  --  (\Lh+2*\dL*\LL,-\hs) 
	--(2*\Lh,-\hs)--(2*\Lh,1.5*\wy)--(2*\Lh+\tWidth,1.5*\wy)--(2*\Lh+\tWidth,-\hs-\tWidth)--(\Lh+\tWidth,-\hs-\tWidth)--(\Lh+\tWidth,-\hd-\tWidth)--(-\tWidth,-\hd-\tWidth)--
	(-\tWidth,-.5*\hd)--	(-\tWidth+2*\wy,1.5*\wy)--	(2*\wy,1.5*\wy)--	(0,-.5*\hd);
	\fill[fill=black] (\Lh+.5*\dL,-\hs) circle (1.5*\dL);

%	\draw[dashed] (0,0)--(2*\Lh,0);

	% draw arrow over paddle
	\draw[<->] (\Lh-.75*\LL,-\hs+1.5*\LL) arc (100:170:\LL);
	
	% draw arrow and ball over left wall
	\draw[<->] (2*\tWidth,-.2*\hd) arc (45:180-45:\LL);
	\fill[fill=black] (0,-.5*\hd) circle (1.5*\dL);
	
	% depth linee
	\node[anchor=west] at (0.05*\hd,-.4*\hd) {$\zWall_1(t)$};
	\node[anchor=east] at (2*\Lh,-.5*\hs) {$\zWall_2(t)$};
	\node[anchor=south] at (.5*\Lh,-\hd) {$\zBed(t)$};

%%	\node[anchor=center] at (.6*\Lh,-.4*\hd) {$\pt(\wx,\wy,t)$};
%	\node[anchor=center] at (.65*\Lh,-.4*\hd) {$\w(\z,t)$};
		\node[anchor=center] at (.65*\Lh,-.4*\hd) {$\zDomain(t)$};
	
	% draw wave
	\begin{scope}
		\clip  (1.4*\wy,-\hs) rectangle (\L,1.5*\wy);
		\draw (0,-.75*\wy) cos +(\wx,.75*\wy)  sin +(\wx,.75*\wy) cos +(\wx,-.75*\wy) sin +(\wx,-.75*\wy) cos +(\wx,\wy) sin +(\wx,1.2*\wy) cos +(\wx,-1.2*\wy) sin +(\wx,-\wy);% cos +(\wx,1.5*\wy); %sin +(\wx,\wy) cos +(\wx,-\wy) sin +(\wx,-\wy) cos +(\wx,\wy);
	\end{scope}
	\draw[dotted] (0,0) (2*\Lh,0);
	\def\wxh{1.5*\lam}
%	\draw[thick,<->] (\wxh,0)--(\wxh,1.4*\wy);
%	\node[anchor=west] at (\wxh,.5*1.4*\wy) {$\etaz(\x,t)$};
	\node[anchor=center] at (.75*\wxh,.5*1.4*\wy) {$\zSurface(t)$};

%	% draw wavemaker flap
%\def\pard{1/8*\L}
%\def\flapFill{gray}
%	\def\parw{0.025*\Lh} % flap width
%	\def\fLen{{(\pard+1.45*\wy)}}%flap length
%	\def\dX{2*\wy}
%	\draw[thick,fill=\flapFill] ({\parw+\dX},-\pard)--++(0,\fLen)--++({-2*\parw},0)--++(0,-\fLen) -- cycle;
%	\draw[ultra thick]  (0,{-2/3*\pard}) --+ ({\dX-\parw},0);
%	\draw[ultra thick]  (0,{-1/6*\pard}) --+ ({\dX-\parw},0);
\end{tikzpicture}%

%%%%%%%%%%%%%%%%%%%%%%%%%%%%%%%%%%%%%%%%%%

%\end{document}
\label{fig:fSketch:z}}\\%
	\subfloat[Prescribed $\zz$-plane.]{%\documentclass{article}
%\usepackage{tikz}
%\usetikzlibrary{patterns}
%\usepackage{amsmath,bm,amsfonts,amssymb}
%
%\newcommand{\fvar}{\bar}
%\newcommand{\etazz}{\fvar y^{\mathrm{s}}}
%
%\newcommand{\xx}{\fvar x}
%\newcommand{\zz}{\fvar z}
%\newcommand{\ww}{\fvar w}
%\newcommand{\hh}{\fvar h}
%
%\n\newcommand{\zDomain}{\mc Z}
%\newcommand{\zzDomain}{\zzvar{\mc Z}}
%\newcommand{\zzzDomain}{\zzzvar{\mc Z}}
%\newcommand{\zBed}{\partial\zDomain\^b}
%\newcommand{\zWall}{\partial\zDomain\^w}
%\newcommand{\zSurface}{\partial\zDomain\^s}
%\newcommand{\zzWall}{\partial\zzDomain\^w}
%\newcommand{\zzBed}{\partial\zzDomain\^b}
%\newcommand{\zzSurface}{\partial\zzDomain\^s}
%\newcommand{\zzzSurface}{\partial\zzzDomain\^s}
%\newcommand{\zBoundary}{\partial\mc Z}
%			
%\begin{document}

%%%%%%%%%%%%%%%%%%%%%%%%%%%%%%%%%%%%%%%%%%

\begin{tikzpicture}[scale=.75]%,font=\large]
	
	% visual
	\def\L{10} % domain length
	\def\lam{5} % wavelength
	\def\hd{1/4*\L}
	\def\Lh{1/2*\L} % gap width/half length
	\def\wy{.07*\Lh} % wave height
	\def\wx{.25*\lam} % wave quarter length
	\def\tWidth{.05*\Lh} % width of surface texture pattern
	
	% crop image
	\clip (-\tWidth,3*\wy) rectangle (\L+\tWidth,-\hd-\tWidth);
	
	% drawin bed
	\draw[thick]  (0,{1.5*\wy})--(0,-\hd)--(2*\Lh,-\hd)--(2*\Lh,{1.5*\wy});
	\fill[pattern=north east lines, pattern color=black]  (0,-\hd)--(2*\Lh,-\hd)--(2*\Lh,1.5*\wy)--(2*\Lh+\tWidth,1.5*\wy)--(2*\Lh+\tWidth,-\hd-\tWidth)--(-\tWidth,-\hd-\tWidth)--(-\tWidth,1.5*\wy)--(0,1.5*\wy);

	\draw[dashed] (0,0)--(2*\Lh,0);

	% depth lines
	\def\wxh{\lam/2}
	\draw[thick,<->] (\wxh,0)--(\wxh,-\hd);
	\node[anchor=west] at (\wxh,-.5*\hd) {$\hh$};
%	\draw[thick,<->] (\wxh,0)--(\wxh,1.3*\wy);
%	\node[anchor=west] at (\wxh,1.3*.5*\wy) {$\etazz(\xx,t)$};
		\node[anchor=base west] at (2.5*\wxh,\wy) {$\zzSurface(t)$};
	
%%	\node[anchor=center] at (\Lh,-.5*\hd) {$\pp(\wxx,\wyy,t)$};
%	\node[anchor=center] at (\Lh,-.4*\hd) {$\ww(\zz,t)$};
	\node[anchor=center] at (\Lh,-.4*\hd) {$\zzDomain(t)$};
	
	% draw wace
	\draw (0,-\wy) cos +(\wx,1.3*\wy)  sin +(\wx,\wy) cos +(\wx,-\wy) sin +(\wx,-\wy) cos +(\wx,.75*\wy) sin +(\wx,.75*\wy) cos +(\wx,-.75*\wy) sin +(\wx,-.75*\wy);% cos +(\wx,\wy); %sin +(\wx,\wy) cos +(\wx,-\wy) sin +(\wx,-\wy) cos +(\wx,\wy);
	\draw[dotted] (0,0) (2*\Lh,0);
	
	\node[anchor=base west] at (-\tWidth,1.75*\wy) {$\zzWall_1=\{\xx\^w_1+\ii\yy\}$};% {$\zzWall_1=\{\xx=\xx\^w_1\}$};
%	\node[anchor=east] at (2*\Lh,-.5*\hd) {$\zzWall_2=\{\xx=\xx_2\^w\}$};
	\node[anchor=south] at (1.25\Lh,-\hd) {$\zzBed=\{\xx-\ii\hh\}$}; %{$\zBed=\{\yy=-\hh\}$};

\end{tikzpicture}%

%%%%%%%%%%%%%%%%%%%%%%%%%%%%%%%%%%%%%%%%%%

%\end{document}
\label{fig:fSketch:zz}}\hfill%
	\subfloat[Rectangular $\zzz$-plane.]{%\documentclass{article}
%\usepackage{tikz}
%\usetikzlibrary{patterns}
%\usepackage{amsmath,bm,amsfonts,amssymb}
%
%\newcommand{\fvar}{\bar}
%\newcommand{\etazz}{\fvar y^{\mathrm{s}}}
%\newcommand{\xx}{{\fvar x}}
%
%\newcommand{\ffvar}[1]{\bar{\bar{#1}}}
%\newcommand{\zzmap}{\ffvar f}
%\newcommand{\xxx}{{\ffvar x}}
%\newcommand{\yyy}{{\ffvar y}}
%\newcommand{\zzz}{{\ffvar z}}
%\newcommand{\www}{\ffvar w}
%\newcommand{\hhh}{{\ffvar h}}
%\newcommand{\etaOfxxx}{\ffvar\eta}
%
%\newcommand{\zDomain}{\mc Z}
%\newcommand{\zzDomain}{\zzvar{\mc Z}}
%\newcommand{\zzzDomain}{\zzzvar{\mc Z}}
%\newcommand{\zBed}{\partial\zDomain\^b}
%\newcommand{\zWall}{\partial\zDomain\^w}
%\newcommand{\zSurface}{\partial\zDomain\^s}
%\newcommand{\zzWall}{\partial\zzDomain\^w}
%\newcommand{\zzBed}{\partial\zzDomain\^b}
%\newcommand{\zzSurface}{\partial\zzDomain\^s}
%\newcommand{\zzzSurface}{\partial\zzzDomain\^s}
%\newcommand{\zBoundary}{\partial\mc Z}
%			
%\begin{document}

%%%%%%%%%%%%%%%%%%%%%%%%%%%%%%%%%%%%%%%%%%

\begin{tikzpicture}[scale=.75]%,font=\large]
	
	% visual
	\def\L{10} % domain length
	\def\lam{5} % wavelength
	\def\hd{1/4*\L}
	\def\Lh{1/2*\L} % gap width/half length
	\def\wy{.07*\Lh} % wave height
	\def\wx{.25*\lam} % wave quarter length
	\def\tWidth{.05*\Lh} % width of surface texture pattern
	
	% crop image
	\clip (-\tWidth,3*\wy) rectangle (\L+\tWidth,-\hd-\tWidth);

	% drawin bed
	\draw[thick]  (0,{1.5*\wy})--(0,-\hd)--(2*\Lh,-\hd)--(2*\Lh,{1.5*\wy});
	\fill[pattern=north east lines, pattern color=black]  (0,-\hd)--(2*\Lh,-\hd)--(2*\Lh,1.5*\wy)--(2*\Lh+\tWidth,1.5*\wy)--(2*\Lh+\tWidth,-\hd-\tWidth)--(-\tWidth,-\hd-\tWidth)--(-\tWidth,1.5*\wy)--(0,1.5*\wy);

	\draw (0,0)--(2*\Lh,0);

	% depth lines
	\def\wxh{\lam/2}
	\draw[thick,<->] (\wxh,0)--(\wxh,-\hd);
	\node[anchor=west] at (\wxh,-.5*\hd) {$\hhh(t)$};
%	\draw[thick,<->] (\wxh,0)--(\wxh,1.3*\wy);
%	\node[anchor=west] at (\wxh,1.3*.5*\wy) {$\etazz(x,t)$};
%	\node[anchor=west] at (\wxh,1.3*.5*\wy) {$\etaOfxxx(\xxx,t)=\Im[\zzmap(\xxx,t)]$};
%	\node[anchor=west] at (\wxh,1.3*.5*\wy) {$\etaOfxxx(\xxx,t)\mapsto\etazz(\xx,t)$};
	\node[anchor=south] at (\Lh,0) {$\zzzSurface = \{\xxx\}$};%{$\zzzSurface = \{\yyy=0\}$};
	
%	\node[anchor=center] at (\Lh,-.5*\hd) {$\www(\zzz,t)$};
	\node[anchor=center] at (\Lh,-.5*\hd) {$\zzzDomain$};

\end{tikzpicture}%

%%%%%%%%%%%%%%%%%%%%%%%%%%%%%%%%%%%%%%%%%%

%\end{document}
\label{fig:fSketch:zzz}}%
	\caption{Sketch of the $\zDomain$, $\zzDomain$ and $\zzzDomain$-domains.
		  }
	\label{fig:fSketch}
\end{figure}

The conformal mapping approach simplifies evaluation of complex  boundaries such as sketched in \autoref{fig:fSketch:z}.
Consider  a moving domain boundary, $\zBoundary(t)$, along which we aim to enforce a kinematic condition of impermeability. 
A general means of doing so is through a fluid particle position $z\^p(t)$  advected with the fluid velocity. 
The condition states that a particle initially located at the boundary will remain there:
\begin{equation}
	z\^p(t) \in  \zBoundary(t) \;\text{ provided }\;z\^p(0) \in  \zBoundary(0).
	\label{eq:kinBC:z}%
\end{equation}
While valid, this formulation is cumbersome to work with directly, motivating the use of conformal mapping
Maps are sought which can project an arbitrarily complex transient domain into a fixed rectangle or strip $\zzzDomain$ (\autoref{fig:fSketch:zzz}) where \eqref{eq:kinBC:z} can be explicitly evaluated. % where  boundary $\zzzBoundary$ consists of fixed straight lines. 

Harmonic maps have the 
characteristic of mutually dependent coordinates in the sense that shifting one will affect all others.
While the bed and wall boundaries are assumed prescribed, 
the free surface evolves dynamically and cannot be predetermined.
Mapping all boundaries simultaneously would therefore require continuous adjustments,  for example using  iteration as in \citet{viotti2014conformal}.
To circumvent the need for iteration, we introduce an intermediate mapping layer
$\zz$ (\autoref{fig:fSketch:zz}) in which all prescribed boundaries are fixed straight lines.
The free surface is then mapped to a straight surface line in a third rectangular $\zzz$-plane (\autoref{fig:fSketch:zzz}).
Impermeability now ensures that any movement on the boundary $\zzBoundary$ remains purely tangential such that boundary points remain on the prescribed contour in accordance with the kinematic condition \eqref{eq:kinBC:z}.

The mapping $\zzz\mapsto\zz\mapsto\z$ is represented through functions $\zmap$ and $\zzmap$,
\begin{equation}
	\z = \zmap(\zz,t),\qquad \zz = \zzmap(\zzz,t),
\end{equation}
and we introduce the conversion throughout of marking functions that take $\zz$ and $\zzz$-coordinate arguments with the same single and double bar notation, respectively. 
This conformal mapping approach is works efficiently because simulation is performed entirely from the $\zzz$-plane, never requiring inverse mapping apart from setting up initial conditions. 
The simulation routine, as well as the mapping of the final solution, is then entirely explicit.

The prescribed mapping function $\zmap(\zz,t)$, describing walls, bathymetry and imposed external motions,  is expressed numerically or analytically prior to simulation. %to be computed in advance of the simulation itself, or provided as an analytical expression. 
Two families of mapping functions, detailed in \autoref{sec:map:SC}, accommodate a broad range of application.
In cases where periodic boundaries replace solid walls, the $\zzz$-plane transforms into a periodic strip.
A third possibility, as noted by  \citet{ruban2005}, is a shore edge type of boundary. This is achieved  by letting $\zmap(\zz,t)$ asymptotically tend towards a single point $ C_\pm(t)$: $\zmap(\zz,t)\to C_\pm(t)$ as $\xx\to\pm\infty$. Shore boundaries have not been tested in the present study.
\\

Fluid velocities are represented using the complex potential  $\w = \phi+\ii\psi$, $\phi$ being the fluid velocity potential and $\psi$ its harmonic conjugate, the stream function. 
By virtue of conformality, these functions transfer directly between planes,
\begin{equation}
	\ww(\zz,t)=\w[\zmap(\zz,t),t], \qquad \www(\zzz,t)=\ww[\zzmap(\zzz,t),t],
	\label{eq:w}
\end{equation}
while still satisfying the Laplace equation.
Differentiation now yields the following key properties:
\begin{equation}
	\begin{aligned}
		\w_\z = \ww_\zz/\zmap_\zz, \quad \w_t = \ww_t - \ww_\zz \zmap_t/\zmap_\zz,\\
		\ww_\zz = \www_\zzz/\zzmap_\zzz, \quad \ww_t = \www_t - \www_\zzz \zzmap_t/\zzmap_\zzz.
	\end{aligned}
	\label{eq:diffprop}
\end{equation}
In anticipation of wall conditions to come, we also introduce an arbitrary 
 pre-determined background potential  $\W(\z,t)$, the total potential then being  $\w+\W$. 
The background potential is relevant only  for transient wall boundaries and vanishes ($\W=0$ ) when walls are stationary or periodic.
A similar background field was introduced by \cite{ducrozet2016_openSourceHOS}; however, our approach differs in that it maintains exactness.
The  background potential is imposed in the $\zz$-plane, where wall boundaries are steady and the mapping predetermined; $\WW(\zz,t)=\W[\zmap(\zz,t),t]$. 
Defined the same way, $\WW$ also obeys the differentiation rules stated above.

\subsection{The kinematic boundary condition}
The kinematic boundary condition \eqref{eq:kinBC:z} can now be reformulated in the  rectangular $\zzz$ plane where evaluation is more convenient.
The position of a fluid particle $\z\^p(t)$ of \eqref{eq:kinBC:z}   maps directly between the different coordinate systems:
\begin{equation}
	\z\^p(t) = \zmap[\zz\^p(t),t], \qquad	\zz\^p(t) = \zzmap[\zzz\^p(t),t].
	\label{eq:zp}
\end{equation}
This leads us to the notion of a particle velocity $\mc U$ as viewed in the $\z$-plane, as well as  \textit{apparent particle velocities} $\UU$ and $\UUU$ as viewed in the intermediate $\zz$  and final $\zzz$-plane, respectively.
Using \eqref{eq:diffprop} and \eqref{eq:zp}, we obtain
\begin{subequations}
	\begin{align}
		\mc U &\equiv	\ddfrac{\z\^p}{t} = \w_\z^* + \W_\z^*,\\
		\UU &\equiv \ddfrac{\zz\^p}{t} =  \frac{\mc U-\zmap_t}{\zmap_\zz}= 	\frac{\ww_\zz^*+\WW_\zz^*}{|\zmap_\zz|^{2}}-\frac{\zmap_t}{\zmap_\zz},\label{eq:UU}\\
		\UUU &\equiv \ddfrac{\zzz\^p}{t} = \frac{\UU-\zzmap_t}{\zzmap_\zzz}= 		\frac{\www_\zzz^*}{ |\zzmap_\zzz|^2   |\zmap_\zz|^2}  +  \frac{1}{\zzmap_\zzz}\rbr{\frac{\WW_\zz^*}{|\zmap_\zz|^2}-\frac{\zmap_t}{\zmap_\zz}-\zzmap_t  },	\label{eq:UUU}
	\end{align}%
\end{subequations}%
asterisk denoting complex conjugation.
In the $\zzz$-plane, kinematic boundary condition  \eqref{eq:kinBC:z}  
simplifies to 
\begin{subequations}
	\begin{alignat}{2}
		\Im\,\UUU(\xxx,t) &= 0,	
		&\quad
		\Im\,\UUU(\xxx-\ii\hhh,t) &= -\hhh_t,	
			\label{eq:ImUUU}
	\intertext{ and, if walls are present along $x=\xxx\_1\^w$ and $\xxx\_2\^w$,}
		\Re\,\UUU(\xxx\_1\^w+\ii \yyy,t) &=0,
		& \quad 
		 \Re\,\UUU(\xxx_2\^w+\ii \yyy,t) &= 0.
		\label{eq:ReUUU} 	
	\end{alignat}%
	\label{eq:BCUUU}%
\end{subequations}%
We further  have that $\zzmap_t$ is real at the bed and  pure imaginary at walls, and that $\zzmap_\zzz$ is real along both bed and walls.

\subsection{Wall boundaries}
The present method employs discrete Fourier transformation, which inherently enforces periodic boundary conditions. 
If walls are prescribed, this property is further utilized to impose impermeability using the method of mirroring;
consider a periodic variable $\fffun$  described along discrete positions $\{\xxx_i\}$ in a vector $[\fffun_1,\fffun_2,\ldots,\fffun_\N]$. 
Combined with Fourier decomposition, the property $\fffun_\xxx=0$  is enforced along $\xxx=\xxx_1\^w=\xxx_1$ and $\xxx_2\^w=\xxx_N$ by extending the vector with its mirror image:
\begin{equation}
	[\fffun_i] \coloneqq  [\fffun_1,\ldots,\fffun_{ \N},\fffun_{ \N-1},\ldots,\fffun_2].
	\label{eq:mirrorWall}
\end{equation}
Here, $\zmap_\zz(\zz,t)$ is  assumed to be real at the lateral boundaries implying no coordinate rotation.
Similar to a padding routine, we mirror the inputs to Fourier transformation and remove the mirrored half following inverse transformation, as illustrated in \autoref{list:CS} in the appendix.
Alternatively, the folded domain can be simulated in its entirety, including the mirror plane upon initiation.

Mirroring effectively satisfies the impermeability condition
 \eqref{eq:ReUUU} as long as the wall boundaries remain stationary. 
If  walls are moving, then \eqref{eq:ReUUU} along with \eqref{eq:UU} defines a condition for the background potential $\WW$, introduced  for this purpose. 
The resulting condition is
\begin{equation}
	\Re\WW_\zz = \Re(\zmap_\zz\zmap_t^*) \quad \text{along } \xxx=\xxx_1\^w \text{ and } \xxx_2\^w.
\end{equation}

\subsection{Simulation}
Next, we consider  the kinematic condition \eqref{eq:ImUUU} along  the free surface and the bed, both of which remain horizontal  in the $\zzz$-plane.
The requirement that the bed remains flat and steady%
\footnote{The requirement $\hh_t = 0$, which simplifies expressions that follow, can always be accommodated by designs of $\zmap(\zz,t)$.}  %#hht
 in the $\zz$-plane implies that $\Im\UU=0$. Consequently, using  \eqref{eq:ImUUU} and the second expression in	\eqref{eq:UUU}, we find that
\begin{equation}
	\qquad\Im\rbr{\frac{\zzmap_t}{\zzmap_\zzz}}_{\yyy=-\ii\hhh} = \hhh_t(t).
	\label{eq:zzzEvolutionBed}%
\end{equation}
This constitutes a condition on the bottom velocity field, which in the $\zzz$-plane, according to \eqref{eq:ImUUU} and \eqref{eq:UUU}, becomes
\begin{subequations}
\begin{align}
	\ppphi_\yyy(\xxx,-\hhh,t) &= \left.\zzmap_\zzz\Im (\WW_\zz-\zmap_\zz \zmap_t^*)   \right|_{\yyy=-\hhh} \equiv \mub(\xxx,t).
	 \label{eq:BCF:Im}
\intertext{
In cases where the bed remains stationary, we simply have $\mub = 0$.
Conversely, at the free surface, the velocity potential is prescribed rather than the position, and we express  \eqref{eq:ImUUU} and \eqref{eq:UUU} accordingly:
}
\Im\rbr{\frac{\zzmap_t}{\zzmap_\zzz}}_{\yyy=0} &= \left. -|\zzmap_\zzz\zmap_\zz|^{-2} \Im\sbr{ \www_\zzz +  \zzmap_\zzz\rbr{\WW_\zz- \zmap_\zz\zmap_t^* } }    \right|_{\yyy=0} \equiv \mus(\xxx,t) . \label{eq:zzzEvolution}  
	\end{align}%
	\label{eq:BCF}%
\end{subequations}%

Our method repeatedly relies on the projection kernels described in \appendixref{sec:mapChalikov}.
Using these, the complex potential $\www$ is projected onto  the surface potential 
$\ppphi\^s=\ppphi(\xxx,0,t) = \Re\www(\xxx,t)$
and the bottom velocity condition \eqref{eq:BCF:Im}. 
In form of the velocity field, the projection is
\begin{align}
	\www_\zzz(\zzz,t) &= \frac{\partial }{\partial \zzz}\CC{\hhh}{\ppphi\^s}(\zzz) -\ii \CC{-\hhh}{\mub}(\zzz+\ii\hhh)
	\nonumber\\&
	= \mc F\^{-1}\sbr{ \rbr{ \kk  \FF(\ppphi\^s) \ee^{-\kk\hhh}
			-  \FF(\mub)} \ii \sech(\kk\hhh)\ee^{-\kk\yyy}  }.
	\label{eq:www}	
\end{align}
The latter expression illustrates the projection kernel, $\mc F$  implying discrete Fourier transformation with wavenumbers $\kk$; see \appendixref{sec:mapChalikov} for details.
A real constant may be added to this expression to superpose an irrotational current to the solution.

Equation \eqref{eq:zzzEvolution} describes  the evolution of the mapping function $\zzmap$  imposed by  surface dynamics.
Since bed coordinates must remain at $\yy=-\hh$ it follows that $\Im{\zzmap_t(\xxx-\ii\hhh,t)}$ is constant. 
According to \appendixref{sec:mapChalikov}, the appropriated mapping is then
\begin{equation}
    \frac{\zzmap_t}{\zzmap_\zzz} = 	\ii  \SS{\hhh}{\mus}(\zzz) +u_0(t),
	\label{eq:zzztMap}%
\end{equation}%
the real function $u_0(t)$ being arbitrary.
The surface values  $\zzmap_t(\xxx)$ are integrated in time to obtain $\zzmap(\xxx,t)$, determining the evolution of the free surface at the next time step.
To avoid numerical issues, it is advisable to march only the vertical surface component
\begin{equation}
\etazzOfxxx(\xxx,t)=\Im\cbr{ \zzmap(\xxx,t)}
\label{eq:etaDef}
\end{equation}
forwards in time and reconstruct  $\zzmap(\xxx,t)$ at the next time level.
Since $\Im{\zzmap(\xxx-\ii\hhh,t)}$ is constant, the projection 
\begin{equation}
	\zzmap(\zzz,t) = \zzz + \ii \SS{\hhh}{\etazzOfxxx}(\zzz) + \xi_0(t),
	\label{eq:zzzMap}
\end{equation}
is appropriate, $\xi_0(t)$ being another arbitrary real function.
Applying the kernel property \eqref{eq:LProps:SH}, we now determine the depth of the $\zzz$-plane as
\begin{equation}
	\hhh =  \hh + \mean{\etazzOfxxx} .
	\label{eq:hhh}%
\end{equation}

The remaining details of the kinematic condition now concerns the real functions $\xi_0(t)$ and $u_0(t)$ in  \eqref{eq:zzzMap} and \eqref{eq:zzztMap}, which must be chosen consistently. % such that the two equations are consistent.
To do so, we consider the horizontal average%
\footnote{defined $\mean{\zzzvar \mu}=  \int_0^{\zzzvar{L}} \zzzvar \mu(\xxx+\ii \yyy)\,\dd\xxx/\zzzvar{L}   =\FF_{\!0}(\zzzvar \mu).  $}
of $\zzmap$ and $\zzmap_t$. Combining \eqref{eq:zzztMap}  and \eqref{eq:zzzMap} and employing the projection kernel definitions, we find
\begin{subequations}
\begin{align}
	\mean{\zzmap} &=  \ii \!\mean{\etazzOfxxx} + \mean{\zzz} + \xi_0(t),	\label{eq:mean_f}\\
	\mean{\zzmap_t}&= \ii \!\mean{\mu} + u_0(t) - \sum_{j=1}^\MM \frac{2 \kk_j}{\sinh^2\kk_j\hhh}  \Im\sbr{ \mc \FF_{\!j}(\etazzOfxxx)\FF_{\!-j}(\mus) }.	\label{eq:mean_ft}%
\end{align}%
\label{eq:mean}%
\end{subequations}%
To maintain a non-drifting frame we set $ \xi_0(t)=0$. Consistency is then maintained
by having the two last terms of \eqref{eq:mean_ft} cancel. % to allow  $\xi_0(t)=0$. 
The easiest way to achieve this in practice is setting 
\begin{equation}
	 u_0(t) =   \Im\mean{    \zzmap_\zzz \SS{\hhh}{\mus} } .
	\label{eq:xi_u0}
\end{equation}
In closed domains, the expression can be disregarded entirely, as the mirror symmetry of \eqref{eq:mirrorWall} ensures zero mean. 
Equations %The reader may now verify consistency between
	\eqref{eq:zzzEvolutionBed}, \eqref{eq:zzzEvolution},	\eqref{eq:zzztMap}, \eqref{eq:etaDef},	\eqref{eq:hhh},	\eqref{eq:mean_ft}, 	\eqref{eq:LProps:S0} and 	\eqref{eq:LProps:SH}
are now consistent and yield the following relationships of conservation:
\begin{align}
	\Im\mean{\zzmap_t} =\mean{\mus} =  \Im\rbr{\frac{\zzmap_t}{\zzmap_\zzz}}_{\yyy=-\hhh} =  \mean{\etazzOfxxx}_t = \hhh_t.
\end{align}
\\

Having now established surface coordinates and velocities, the final step is to impose the dynamic boundary condition.
In the  $\z$-plane, this is given the Bernoulli equation 
\begin{equation}
	\phi_t + \Re\W_t+ \frac12|\w_\z+\W_\z|^2+g\y + p/\rho=0 
	\label{eq:bernoulli}
\end{equation}
evaluated along  $\zSurface$ where $p=0$.
Using \eqref{eq:diffprop}, it is straightforward to rewrite \eqref{eq:bernoulli} in terms of $\zzz$-variables:
\begin{align}
		\ppphi_t = \Re\sbr{\frac{\www_\zzz}{\zzmap_\zzz}\rbr{\zzmap_t+\frac{\zmap_t}{\zmap_\zz}}  -  \WW_t+\WW_\zz\frac{\zmap_t}{\zmap_\zz}} - \frac12\br|{\frac{\www_\zzz}{\zmap_\zz \zzmap_\zzz}+\frac{\WW_\zz}{\zmap_\zz}}|^2 - g\y %\quad\text{along }\yyy=0
	\label{eq:BC_dyn_zzz}
\end{align}
evaluated along $\yyy=0$ with coordinates mapped accordingly. 
he numerical examples presented are solved using a dynamic time-stepping ODE solver, which integrates either the physical or Fourier components of $\ppphi_t\S$ and $\etazzOfxxx_t$.%
\footnote{
Marching the Fourier variables in time requires fewer Fourier transformations but the ODE solver often responds better to physical variables.
} 
Surface tension can be added to \eqref{eq:BC_dyn_zzz} without difficulty.
This completes the simulated part of the conformal mapping model.
\\

\autoref{list:basic} provides an excerpt of MATLAB code illustrating how the model may be implementation practice. 
With stationary beds, each time step requires seven Fourier transformations, with an additional four needed for the stabilisation procedure described later. 
When modelling transient beds, three more transformations are required.
In the special case of a flat bed, where $\zmap(\zz,t) = \zz$, the model simplifies to a compact version of the model by  \citet{chalikov2005modeling}.
When extended to transient bathymetries and infinite boundaries, the model aligns conceptually with that of \citet{ruban2005}. %, albeit with a different formalism.

\renewcommand{\^}{\slashHat} % earlier redefinition of \^ can cause problems in listing in some documentclass environments 
\begin{lstlisting}[label=list:basic,caption={Implementation example (MATLAB syntax) assuming stationary prescribed map $\zmap(\zz)$ and periodic boundary conditions. Repeated characters replace the single and double bar notation. Anti-aliasing neglected.} ]
	FFTeta	= fft(eta);
	hhh     = hh + mean(eta);                   % eq 16
	C       = 2./(1+exp(2*k*hhh));              % eq A.1a
	S       = 2./(1-exp(2*k*hhh)); S(1) = 1;    % eq A.1b 
	zz      = xxx + 1i*ifft(S.*FFTeta);         % eq 15
	fff_zzz = 1 - ifft(S.*k.*FFTeta);           % eq 15
	J       = abs(ff_zz(zz).*fff_zzz).^(-2);    % double inverse Jacobian
	www_zzz = ifft(C.*1i.*k.*fft(ppphi));       % eq 12
	mu_s    = -J.*imag(www_zzz);                % eq 11b
	ftfz    = 1i*ifft(S.*fft(mu_s));            % eq 13 (ftfz = fff_t/fff_zzz)
	ftfz    = ftfz - mean(real(ftfz.*fff_zzz)); % eq 18 (removes real mean from fff_t)
	eta_t   = imag(ftfz.*fff_zzz);              % eq 14
	ppphi_t = real(www_zzz.*ftfz) - .5*J.*abs(www_zzz).^2 - g*imag(ff(zz)); % eq 21
\end{lstlisting}
\renewcommand{\^}[1]{^\mathrm{#1}}

\subsection{Pressure}
Pressure can  be computed diagnostically during post-processing using the original Bernoulli equation	\eqref{eq:bernoulli}
which provides both the unmapped potential time derivative $\phi_t$ at the surface, and the pressure within the fluid. 
As before, the conjugate of the potential time derivative corresponds to the time derivative of the stream function, which remains constant at the impermeable bed. 
The appropriate projection kernel is in this case  $\mapC_\hhh$ leading to 
\begin{equation}
	\zzzvar p(\zzz,t)/\rho = - \CC{\hhh}{\ppphi_t\S}(\zzz)      - \frac12\br|{\frac{\www_\zzz}{\zmap_\zz \zzmap_\zzz}+\frac{\WW_\zz}{\zmap_\zz}}|^2 - g\y.
	\label{eq:pressure}
\end{equation}
Here, $\ppphi_t\S$ equals the last two terms of \eqref{eq:BC_dyn_zzz} (or equivalently \eqref{eq:pressure}) evaluated at the surface $\yyy=0$.
The flow potential and stream function are obtained directly from  \eqref{eq:www} through straightforward analytical integration, taking case to make the zero-mode proportional to $\zzz$. 

An example of a reconstructed velocity and pressure field, derived from surface variables using equations  \eqref{eq:www},  \eqref{eq:zzzMap} and \eqref{eq:pressure}  is presented in  \autoref{fig:flap45Simulation}.
Additional examples of velocity potentials are shown in \autoref{fig:tsunami:field}.

\subsection{A model hybrid}
A simpler alternative to the method presented thus far is a
conformal mapping extension of the conversional (HOS) scheme, which served as a prototype for the present work.
Although less precise, his approach is also less abstract and therefore worth outlining.
In the prototype, simulations are conducted directly from the $\zz$-plane, assuming a prescribed mapping function $\zmap(\zz)$. The mapping provides the rectangular domain for which the conventional  HOS scheme is designed. 
Defining the surface potential  $\pphi\S(\xx,t)=\pphi[\xx,\etazz(\xx,t),t]$, where  $\etazz(\xx,t)$ describes the surface such that $\xx+\ii \etazz(\xx,t)\in\zzSurface$,   we obtain the governing equations
\begin{subequations}
	\begin{alignat}{2}
		\etazz_t &=	 |\zmap_\zz|^{-2} \sbr{\pphi\S_\xx\etazz_\xx-\rbr{1+\rbr{\etazz_\xx}^2}\pphi_\yy} \quad&&\text{along }\yy=\etazz(\xx,t),\\
		\pphi\S_t &= - 
		\frac12 |\zmap_\zz|^{-2} \sbr{\rbr{\pphi\S_\xx}^2-\rbr{1+\rbr{\etazz_\xx}^2}\pphi_\yy^2} - g\Im\zmap\quad&&\text{along }\yy=\etazz(\xx,t), \\
		\pphi_\yy &= 0\qquad &&\text{along }\yy=-\hh,\label{eq:impermeability:bed}\\
		\pphi_\xx &= 0 &&\text{along }\xx=\xx_j\^w \;\text{(optional)}.\label{eq:impermeability:wall}
	\end{alignat}%
	\label{eq:prototype}%
\end{subequations}%
Vertical surface velocity  $\pphi_\yy[\xx,\etazz(\xx,t),t]$ is computed using the conventional HOS scheme, employing Taylor and Stokes expansions.
Moving boundaries can also be represented by a background potential, here omitted.

\subsection{Stabilisation, anti-aliasing and beach damping}
As with other spectral methods, stabilisation measures are necessary to suppress the buildup of high-frequency noise.
Following \citet{chalikov2005modeling}, 
we adopt the modal damping approach
	\begin{align}
	\etazzOfxxx_t &\coloneqq	\FFInv\sbr{ \FF(\etazzOfxxx_t) -\hat \nu \FF(\eta)}, &
		\ppphi\S_t &\coloneqq 	\FFInv\sbr{  \FF(\ppphi\S_t) -\hat \nu	\FF(\ppphi\S) }.
\label{eq:damping}%
	\end{align}%
As in \citet{chalikov2005modeling}, the modal damping coefficient is modelled 
\begin{equation}
		\hat\nu(k) = 	r\, \sqrt{\frac{2\pi g}{\zzzvar{L}}} \rbr{\frac{\max(|k|-\kd,0)}{\kMax-\kd}}^2,
		\label{eq:damping_nu}
	\end{equation}
	where $\kd$ is an intermediate wavenumber, $\kMax\approx \pi \N/\zzzvar{L}$ is the largest wave number and $\zzzvar{L}$ is the horizontal length of the $\zzzDomain$-domain.
	This formulation enforces no damping up to the intermediate wavenumber, after which damping increases quadratically up to a fixed value.
	Typical  choses are $\kd=0.5\,\kMax$ and $r=0.1$ or $0.25$.

\begin{figure}[H]
		\centering
		\subfloat[Non-breaking wave without stabilisation, $ka = 0.15$.]{\includegraphics[width=.5\columnwidth]{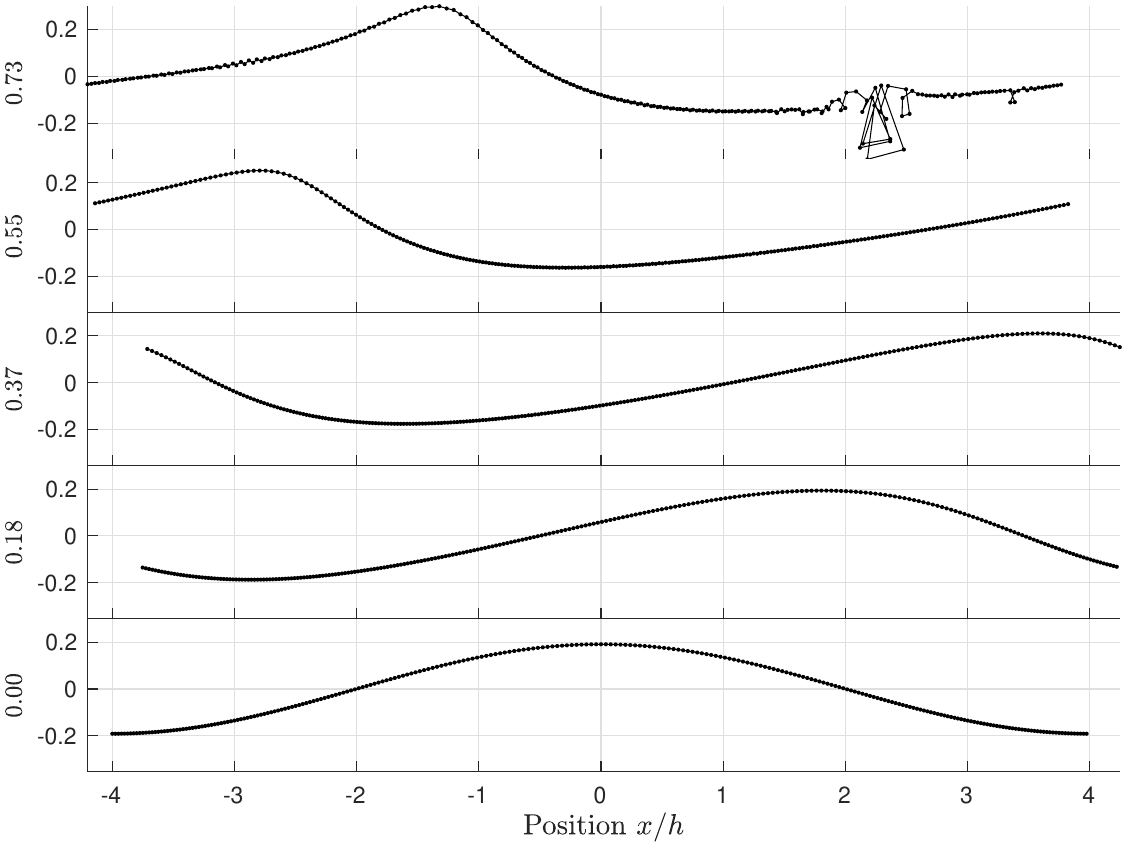}\label{fig:nuStudy:nonBreakingCrash}}%
		\subfloat[Wave breaking suppressed by damping; $ka = 0.20$, $\kd=0.25\,\kMax$, $\rDamping = 0.10$.]{\includegraphics[width=.5\columnwidth]{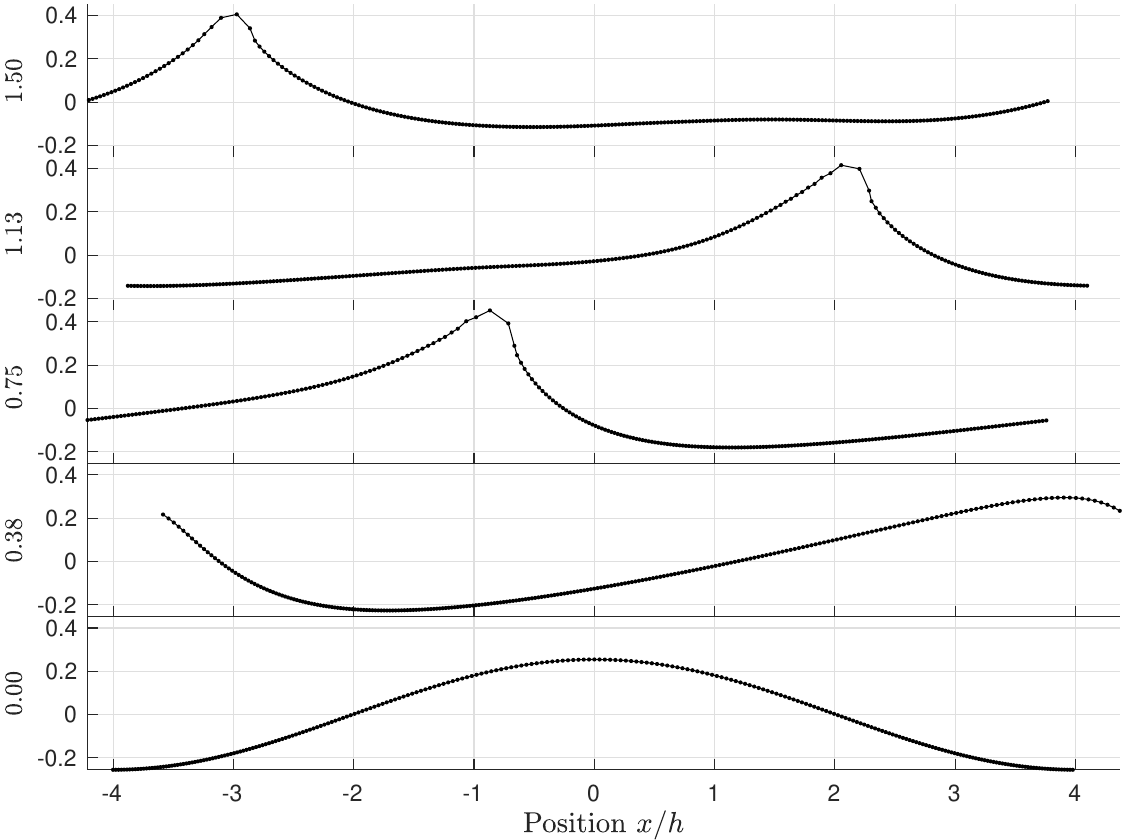}\label{fig:nuStudy:breakingStable}}%
		\\
		\subfloat[Strong wave breaking suppressed by wide-banded damping; $ka = 0.25$, $\kd=0.10\,\kMax$, $\rDamping = 0.10$. ]{\includegraphics[width=.5\columnwidth]{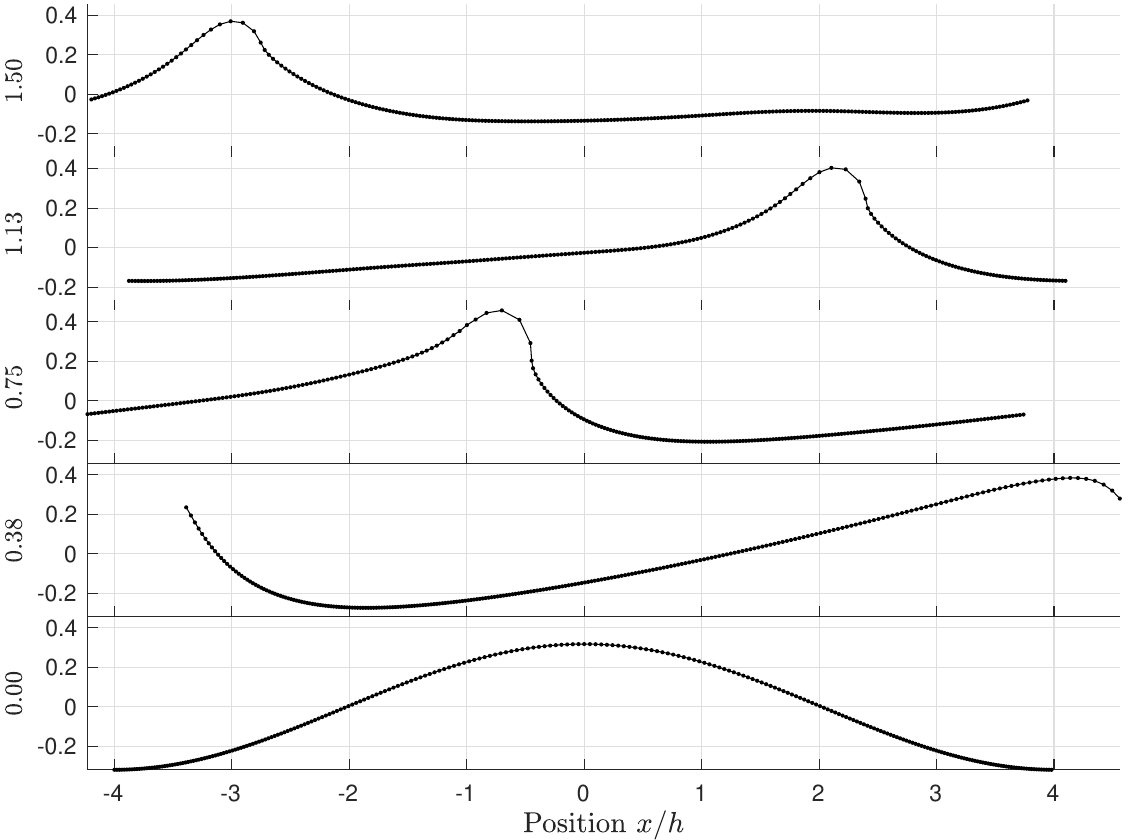}\label{fig:nuStudy:strongBreakingWideBand}}%
		\subfloat[Strong wave breaking suppressed by high-intensity damping; $ka = 0.25$, $\kd=0.25\,\kMax$, $\rDamping = 0.50$.]{\includegraphics[width=.5\columnwidth]{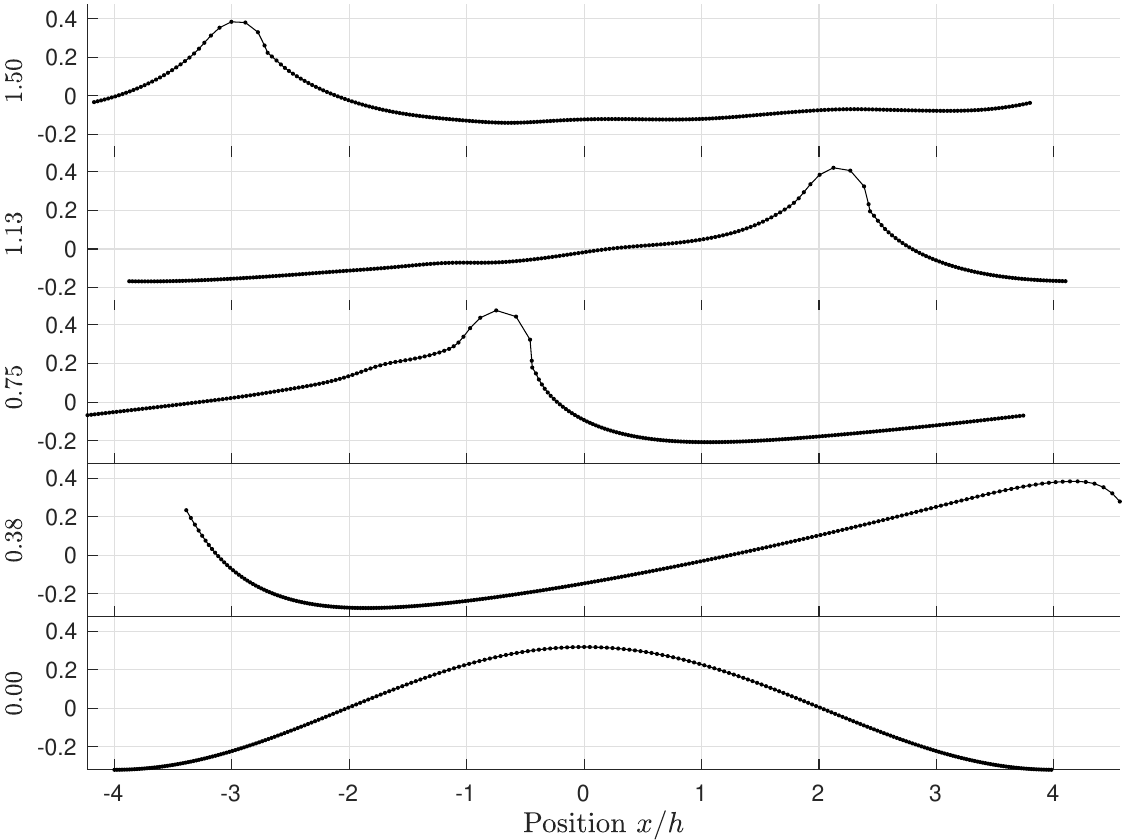}\label{fig:nuStudy:strongBreakingIntense}}%
		\caption{
			\Rone{
				Impact of excessive  stabilisation coefficients on wave dynamics. All examples are initiated with a single-period linear wave at water depth $kh=\pi/4$. Very slight damping is sufficient to stabilise the case in panel \protect\subref{fig:nuStudy:nonBreakingCrash}.}
		}
		\label{fig:nuStudy}
	\end{figure}

\Rone{
	To explore the effects of modal damping, we compare passive stabilization with excessive damping.
Without damping,  high-wavenumber energy accumulates over time, eventually leading to numerical instability, as shown in  \autoref{fig:nuStudy:nonBreakingCrash}.
Only a minimal amount of damping ($\kd=0.50\,\kMax$ and $\rDamping=0.025$) is sufficient to permanently stabilise this case. 
This damping occurs at a scale and magnitude that do not affect wave dynamics.  
In contrast, the other panels of \autoref{fig:nuStudy} illustrate cases where  damping extends into active wave scales:
In panel \subref{fig:nuStudy:breakingStable}, damping is intensified and extended to the scale of the wave front, preventing overtopping and suppressing wave breaking, thus avoiding simulation failure. 
Panels \subref{fig:nuStudy:strongBreakingWideBand} and  \subref{fig:nuStudy:strongBreakingIntense} show an even larger wave, stabilised in panel \subref{fig:nuStudy:strongBreakingWideBand}  by broad-banded damping  (small $\kd$), and in panel \subref{fig:nuStudy:strongBreakingIntense} by high-intensity damping (large $\rDamping$).
The former causes a broader smearing of the wave crest, while an artificial crest `bubble' forms in the latter.
The bubble's length scale is determined by $\kd$ and  resolution, and its formation appears to generate some secondary waves at similar scales. 
While \eqref{eq:damping_nu} effectively maintains numerical stability, it does not constitute a physically realistic wave-breaking model.
Developing a dedicated wave breaking model remains an important direction for future work.
}
\\

Anti-aliasing measures are also a common feature of spectral methods,
addressing resolution errors that arise when nonlinear operations are performed on discrete variables in physical space.
This causes high-wavenumber energy to fold back onto lower wavenumbers.
To mitigate the issue, anti-aliasing involves padding inputs to the inverse Fourier transform, ensuring sufficient resolution in physical space to prevent energy folding  \citep{west1987_originalHOS}. 
The padding that absorbed the trickle down of energy is then removed at the next Fourier transformation.
Since the conformal mapping model involves operations of division we cannot formally eliminate aliasing errors. 
However, \citet{chalikov2005modeling} observed that 
simulations become insensitive to additional padding beyond cubic nonlinearity (i.e., adding zero-modes above wavenumber $3\,\kMax$).
Furthermore, it is the authors experience that aliasing errors rarely impact  short-duration simulations  like those presented here.
Nevertheless, sensitivity should be checked, and anti-aliasing measures included when studying longer wave field evolutions.
\\

Wave absorption is also often needed when simulating closed environments such as wave tanks.
This is usually achieved by including a `numerical beach' into the model, which can either be intended to reproduce the reflection of an actual beach, or just to emulate a radiating boundary.
An approximate variant of the absorption layer method of \citet{bonnefoy2010} is implemented in the author's code,
setting
\begin{align}
	\ppphi_t &\coloneq  \ppphi_t - \zzzvar\nu(\xxx) \y\^s_t,  &  
	\y\^s_t &\approx \Im \rbr{ \zmap_t + \zmap_\zz\zzmap_t }, & 
	\nu & =  \nu_0 u^2(3-2u);\; u = \max(\x-\x\_b,0)/L\_b,
	\label{eq:beach}
\end{align}
absorption intensity $\nu_0$ and beach length $L\_b$ suitably chosen.%^

\section{The mapping function $\zmap(\zz,t)$}
\label{sec:map}
The intermediate mapping function $\zmap(\zz,t)$ maps the bathymetry to a horizontal line at constant  depth $\zz=\xx-\ii\hh$, and  walls, if present, onto stationary vertical lines $\zz = \xx_1\^w+\ii \yy$ and $\zz = \xx_2\^w+\ii \yy$.
Whenever permissible, we also map the still water line $\z=\x$ onto $\zz=\xx$ such that the x-axis of the two planes align. 
This alignment is beneficial for orientation and potentially also stability, but is not a requirement.
We will in this section explore two families of such map functions and also demonstrate the mapping of a piston wavemaker.

\subsection{An analytical bathymetry step}
\label{sec:map:SC}

\begin{figure}[H]%
\subfloat[$\zz$-plane]{\includegraphics[width=.48\columnwidth]{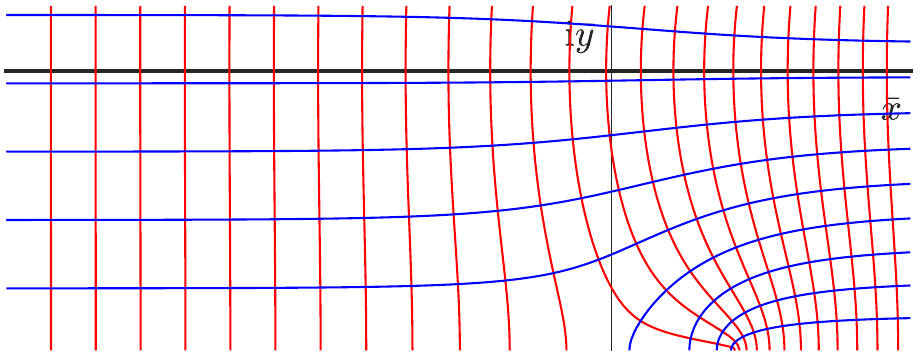}}%
\hfill
\subfloat[$\z$-plane]{\includegraphics[width=.48\columnwidth]{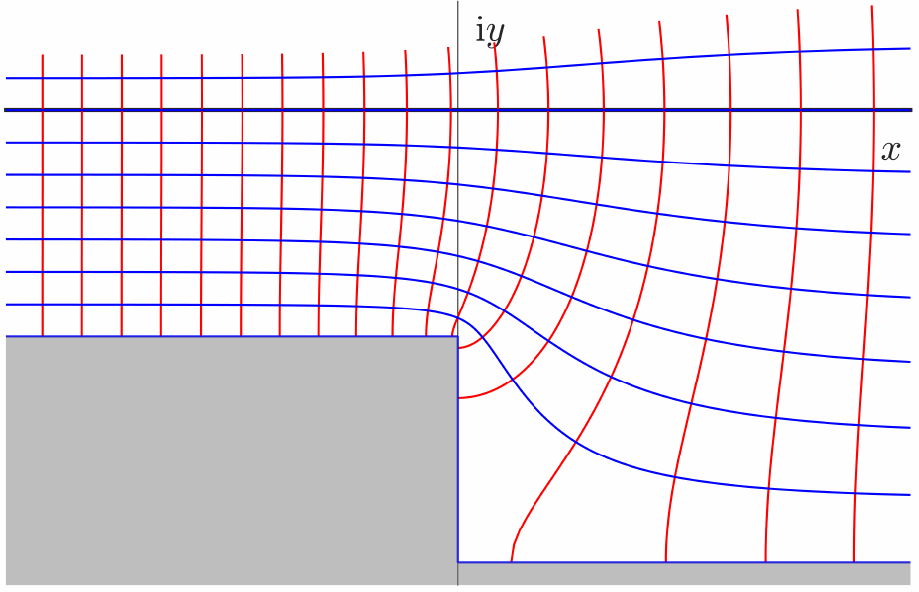}}%
\caption{A step mapped with adjusted Schwarz--Christoffel transform \eqref{eq:map_logstrip}}%
\label{fig:Mei_step}%
\end{figure}

The Schwartz--Christoffel theorem \citep[e.g.,][\SSS10.2]{milneThomson1962theoreticalHydrodynamics} is a powerful tool  constructing mappings with straight-line boundaries and sharp angles. 
The  theorem provides the derivative $\zmap_\zz$ of a function that 
bends the bathymetry line abruptly with prescribed angles at prescribed locations.
When applied directly to a rectangular step, the theorem yields $\zmap_\zz\propto\sqrt{\zz+1}/\sqrt{\zz-1}$,
which can be integrated to match a step bathymetry transition from a shallower depth $h_1$ and deeper depth $h_2$:
\begin{equation}
\zmap(\zz) = \frac{h_2-h_1}{\pi}\rbr{\sqrt{\zz+1}\sqrt{\zz-1}+\arccosh\zz }-\ii \h_2, \quad \Im\zz\geq0.
\label{eq:SCPrimitive}
\end{equation}
However, this map curves along the  real axis  and continues to do so  far away from the step.
To achieve a better mapping---one that remains flat along $\y=0$ through compression and quickly resumes a Cartesian shape---we turn to the textbook example of flow in an abruptly expanding channel 
 \citep[\SSS10.7]{milneThomson1962theoreticalHydrodynamics}. 
The trick of this solution is modelling the flow field using a source placed at infinity, 
and we hare regard this flow field as itself being a coordinate map, yielding
\begin{equation}
\zmap_\zz = \frac{\h\_1}{\pi \tau},\quad
\tau = \sqrt{\frac{ \exp(\zz)+c^2}{ \exp(\zz)+1} },
 \label{eq:map_dlogstrip}%
\end{equation}%
with constant $c = \h\_2/\h\_1$. 
Remarkably, the map is also analytically integrable:\footnote{
The correct logarithm branches must be chosen for the transition into $\y>0$ and so we introduce $\ln^+$ whose branch cut runs along $0$ and $2\pi$. }
	\begin{equation}
		 \zmap(\zz,t) = -\ii \h\_1 +\frac{\h\_2}{\pi}\left[ 
		\frac1c \ln^+\frac{\tau-c}{\tau+c} - \ln\frac{\tau-1}{\tau+1}
		\right]\label{eq:map_logstrip}.
	\end{equation}%
The area from the still water level to the floor is thus mapped into the logarithmic strip 	$-\pi\leq\yy\leq0$ and  mirrored for $0<\yy<\pi$. Consequently, we have $\hh=\pi$. 
An example is shown in \autoref{fig:Mei_step}.
Notably, the still water line $\y=0$ aligns with  $\yy=0$ in the $\zz$-plane and that coordinates are compressed on the shallow side.
The direction of the step is reversed by flipping the abscissa
$z=-\zmap(-\zz)$, and we can of course map sequences of  steps using \eqref{eq:map_logstrip}
provided these are far enough apart to avoid curvature at the seams.

\subsection{Compressed Schwartz--Christoffel maps using numerical integration}
\label{sec:map:SCnum}
When analytical integration fails, numerical integration provides a straightforward alternative
\begin{equation*}
	\zmap(\zz+\Delta \zz)\simeq\zmap(\zz) +  \zmap_\zz(\zz+\Delta \zz/2)\,\Delta\zz
\end{equation*}
where we are free to choose the integration path that best suits our mapping. 
Doing so, expression \eqref{eq:map_dlogstrip} can be generalised to accommodate an arbitrary number  of  angled planes with the expression
\begin{equation}
	\zmap_\zz(\zz,t) =\frac{\h_{1}}{\pi}  \prod_{j=1}^{\NSC} 
	\sbr{\exp(\zz-\xxjSC)+1}^{-\thSC_j/\pi}; \quad \sum_{j=1}^{\NSC} \thSC_j = 0,   % 	\rbr{\ee^{\zz-\xxjSC}+1}^{-\thSC_j/\pi}.
	\label{eq:SCnumMultiStep}
\end{equation}
where outer angles $\{\thSC_j\}$ are as observed in the $\z$-plane, located at coordinates mapped from 
$\zz=\xxjSC-\ii\pi$. % (and $\xxjSC+\ii\pi$).
A corner location will be a root of this expression when the corner is expanding ($\thSC_j<0$) and a singularity when the corner is contracting ($\thSC_j>0$). 
Expression \eqref{eq:SCnumMultiStep} can also be written in the form
$ 	\zmap_\zz(\zz,t) =\frac{\h_{1}}{\pi}  \prod_{j=1}^{\NSC} \rbr{\ee^{\zz}+\ee^{\xxjSC}}^{\thSC_j/\pi}$ 
which more closely resembles the original  Schwartz--Christoffel theorem.
\\

\begin{figure}[H]%
	\subfloat[$\zz$-plane]{\includegraphics[width=.48\columnwidth]{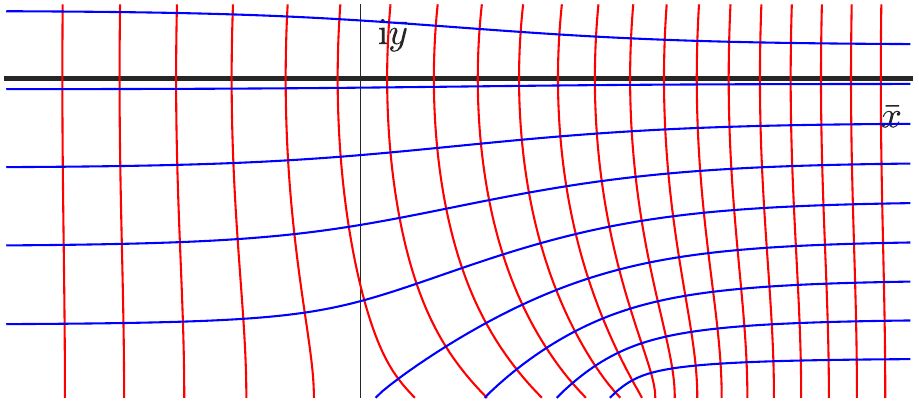}}%
	\hfill
	\subfloat[$\z$-plane]{\includegraphics[width=.48\columnwidth]{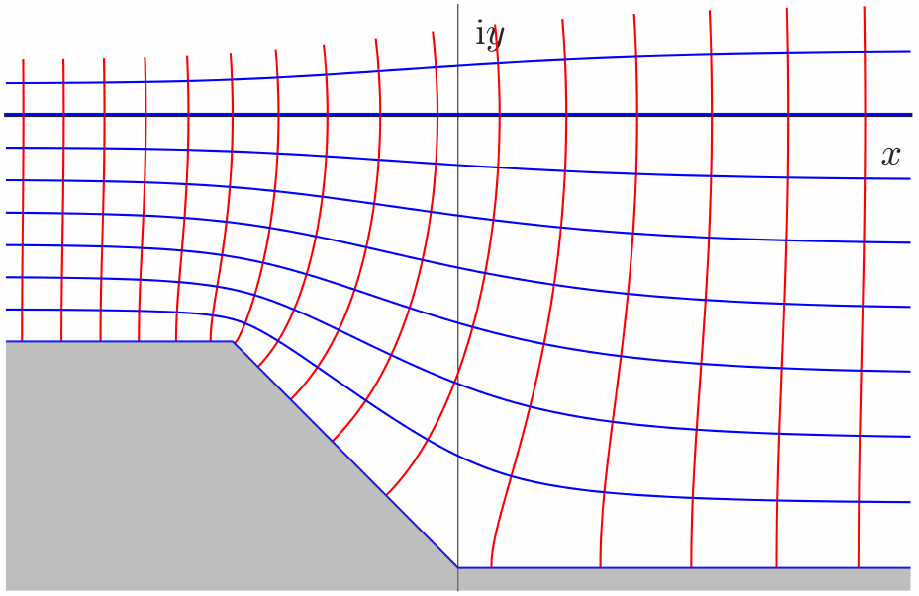}}%
	\caption{Numerical integration of \eqref{eq:SCnumSingleStep} with $\thSC=-\pi/4$, $\h_1 = 0.5$, $h_2=1.0$.	}%
	\label{fig:SCnumStep45deg}%
\end{figure}

\begin{figure}[H]
	\subfloat[{
	$\thSC_j=\big[\frac{\pi}2,\frac{\pi}2,-\frac{\pi}2,-\frac{\pi}2\big]$},\newline {$\xxjSC	=[0.0 ,   0.1 ,   1.1,   1.6]$.%
		}]{\includegraphics[width=.48\columnwidth]{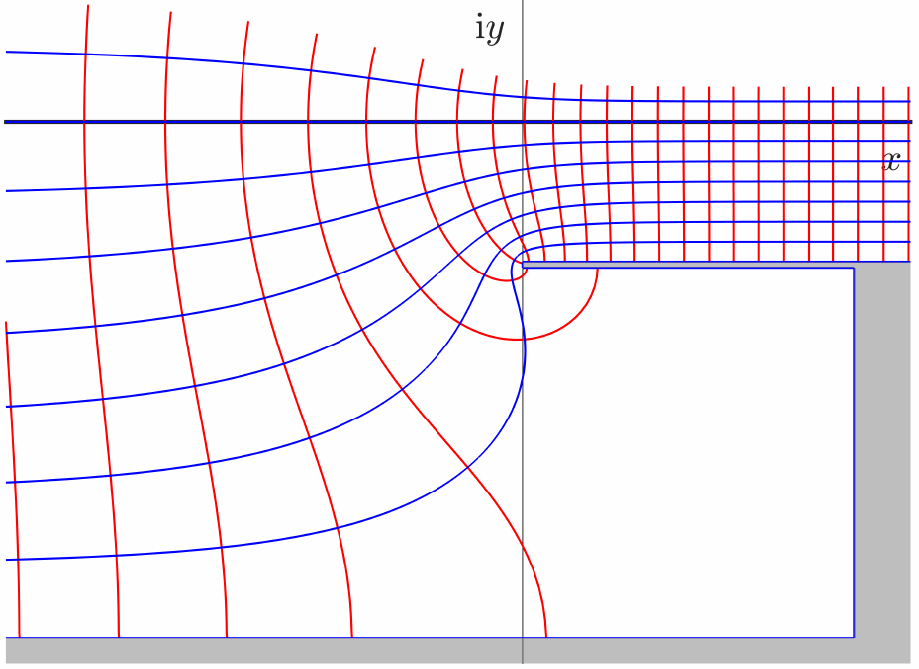}\label{fig:example:flap:0deg}}%
	\hfill
	\subfloat[{%
	$\thSC_j={\big[\frac{\pi}2,\frac{\pi}3,-\frac\pi2,-\frac\pi2,\frac{\pi}6\big]}$,\newline  $\xxjSC=[0, 0.2,2.2,3.2,8.0]$.%
	}]{\includegraphics[width=.48\columnwidth]{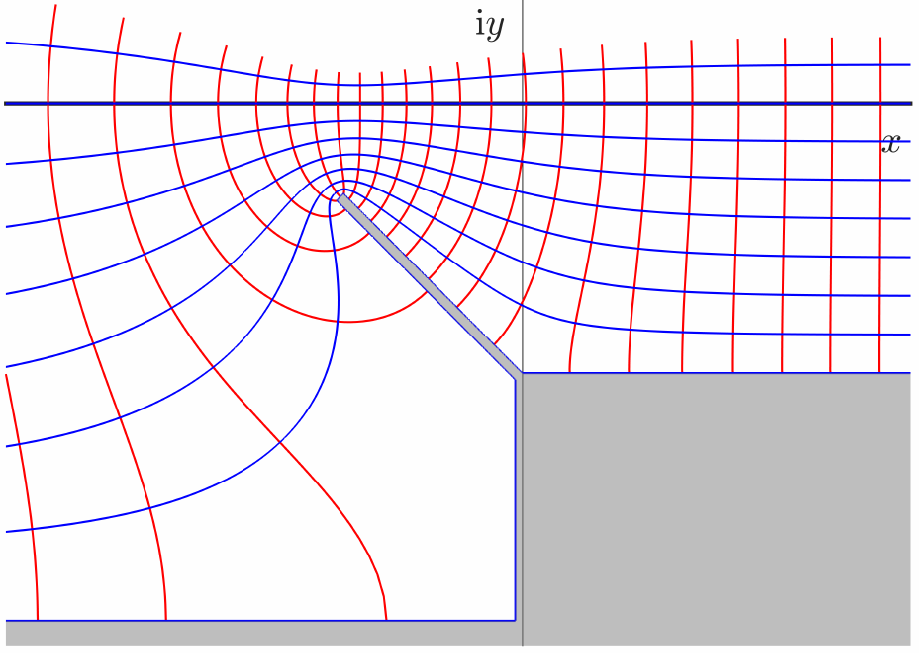}\label{fig:example:flap:45deg}}%
	\caption{Example maps in $\z$-plane constructed using \eqref{eq:SCnumMultiStep}. 
	}
	\label{fig:example:flap}
\end{figure}

\begin{figure}[H]
	\centering
	\includegraphics[width=1\columnwidth,align=t]{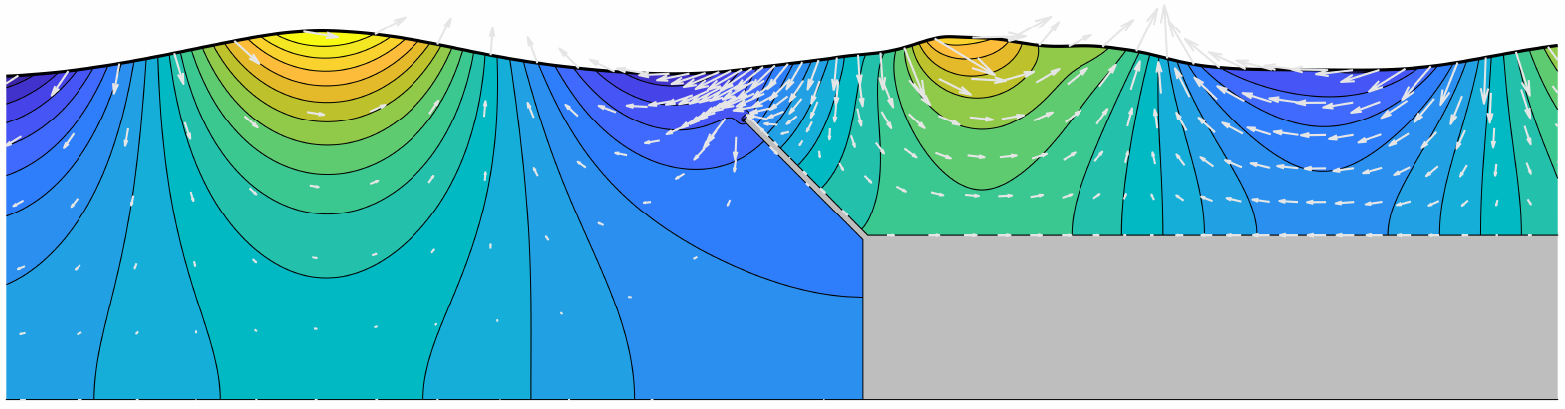}%
	\caption{
		Simulation snapshot showing a regular wave interacting with the geometry in  \autoref{fig:example:flap:45deg}.
		he wave is visibly disturbed by the protruding structure. 
		Contours represent dynamic pressure  \eqref{eq:pressure}, and arrows indicate the velocity field. 
		Arrows near the protruding tip have been omitted for clarity. Initial conditions: wave packet with steepness  $kH/2=0.15$ and water depth $kh_1=3.0$. 
	}
	\label{fig:flap45Simulation}
\end{figure}

The asymptotic limits of $\zmap_\zz(\zz,t)$ are  
\begin{equation}
	\lim_{\xx\to-\infty} 	\zmap_\zz(\zz,t) =  h_1/\pi \quad \text{and}\quad \lim_{\xx\to+\infty} 	\zmap_\zz(\zz,t) = h_\NSC/\pi,
\label{eq:SC:asymptote}
\end{equation}
 $h_1$ and 
\begin{equation}
h_\NSC = h_1 \exp\rbr{ \sum_{j=1}^{\NSC}\frac{\xxjSC \thSC_j}{\pi}}
\label{eq:SC:hN}
\end{equation}
respectively being the water depths found to the left and right of the combined bundle of corners.
We can in other words a priori select $\{\xxjSC\}$ and $\{\thSC_j\}$ to get the desired set of slopes and the desired 
outer water depths. 
For instance, a bathymetry with a single (outer) slope angle $\thSC$, transitioning from depth $h_1$ to $h_2$, is obtained by numerically integrating
\begin{equation}
\zmap_\zz(\zz,t) =\frac{\h_2}{\pi} \rbr{\frac{\exp\zz+1}{\exp\zz + (\h_2/\h_1)^{-\pi/\thSC}}}^{-\thSC/\pi}.
\label{eq:SCnumSingleStep}
\end{equation}
Of course, setting $\thSC=-\pi/2$ recovers \eqref{eq:map_dlogstrip}.
An example of a slanting step is shown in \autoref{fig:SCnumStep45deg} with $\thSC=-\pi/4$.
If additional bathymetry angles are required at prescribed water depths, iteration on $\{\xxjSC\}$ is necessary.
Examples of such geometries are presented in \autoref{fig:example:flap},  showing  two slender protruding structures integrated using \eqref{eq:SCnumMultiStep}. 
The pressure and velocity field for a wave passing above the latter geometry is included in \autoref{fig:flap45Simulation}.
We note the peculiar feature of potential flows, that singularities of pressure and velocity appear at expanding corners, whereas contracting corners exhibit stagnation points. 
This poses no significant challenge computationally, and one can, if preferred, round the corners slightly by intentionally setting $\hh<\pi$.

\subsection{Arbitrary smooth bathymetries}
\newcommand{\HH}{\zzvar H}
The projection kernels  $\mapC$ and $\mapS$ in  \appendixref{sec:mapChalikov}   serve as fundamental components of  the present model.
Similar to the free surface,
these kernels allow for mapping of the bathymetry, provided it is sufficient smooth. 
This is achieved by introducing a $\z$-plane  depth function $H(\x,t)$ defined as $\zBed(t) = \{\x -\ii H(\x,t)\}$,
along with  an equivalent $\zz$-plane function $\HH(\xx,t)$ that takes the same values as $H(\x,t)$ but with $\xx$ as argument.
The mapping is then expressed as
\begin{equation}
	\zmap(\xx,t) = \zz +\ii\hh -\ii \SS{-\hh}{\HH}(\zz+\ii\hh);\quad \hh=\mean{\HH}.
	\label{eq:HH}%
\end{equation}
A quick iteration is necessary to precisely match a prescribed bathymetry because $\HH$ is function of  $\xx$ rather than $\x$.
Since only the horizontal coordinates differ, a fixed-point iteration is usually sufficient---%
an example  is provided in \autoref{list:iterations} and demonstrated in \autoref{fig:HHmap_examples}.
The method performs well with smooth surfaces (\autoref{fig:HHmap_examples:wavey}) but fails when sharp angles are present (\autoref{fig:HHmap_examples:step}).
Fortunately, for such cases, the approach outlined in  \autoref{sec:map:SCnum} is highly effective. 

The iterations in \autoref{list:iterations} can also be adopted to define the initial conditions of $\etazzOfxxx$ and $\ppphi\S$ in the $\zzz$-plane.% prior to simulation.

\begin{lstlisting}[basicstyle=\ttfamily\small,label=list:iterations,caption={Implementation example (MATLAB syntax) of iteration loop that determines $\HH(\xx)$ precisely according to a prescribed bathymetry $H(\x)$. The full map is then given by \eqref{eq:HH}.
	The function \texttt{convKernel} is found in \autoref{list:CS}. Function \texttt{interp1} performs a linear interpolation.
}]
HH = H; xx = x;
for i = 1:n
	  xMap = xx + imag(convKernel('S',-mean(HH),HH,xx,0));
	  HH   = interp1(x,H,xMap);
end
\end{lstlisting}

\begin{figure}[H]
	\centering
	\subfloat[No iteration, setting $\HH=H$.]{\includegraphics[width=.75\columnwidth]{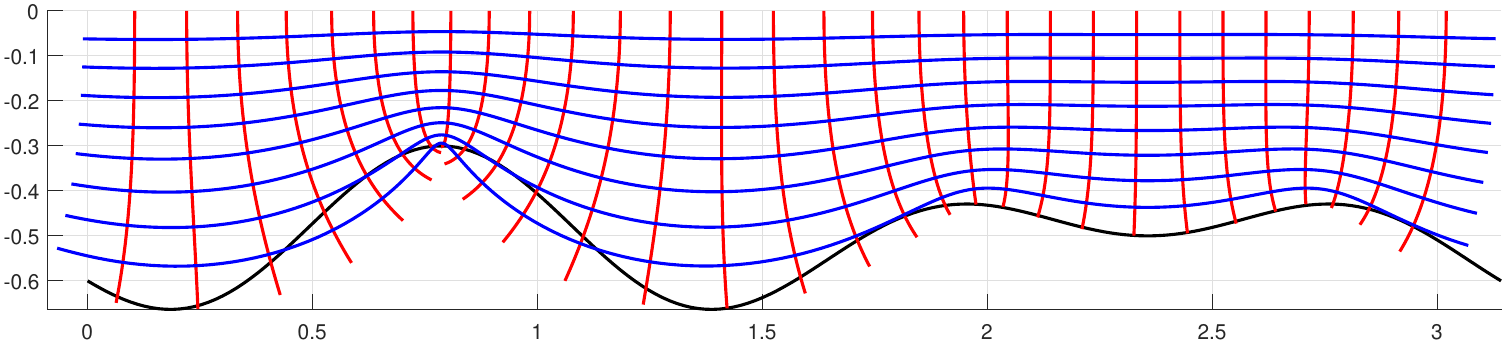}\label{fig:HHmap_examples:wavey0}}\\%
	\subfloat[After a few iterations.]{\includegraphics[width=.75\columnwidth]{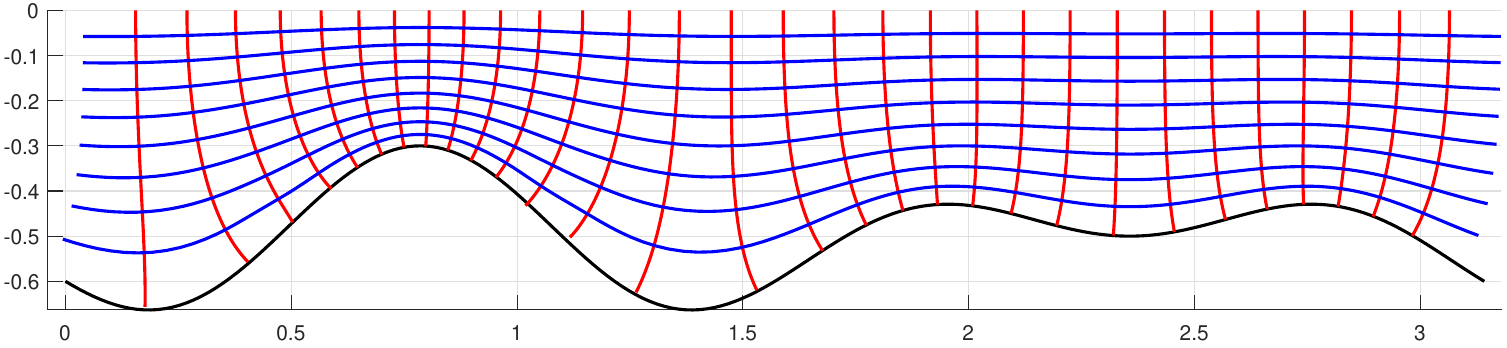}\label{fig:HHmap_examples:wavey}}\\%
	\subfloat[Diverging example: A step.]{\includegraphics[width=.75\columnwidth]{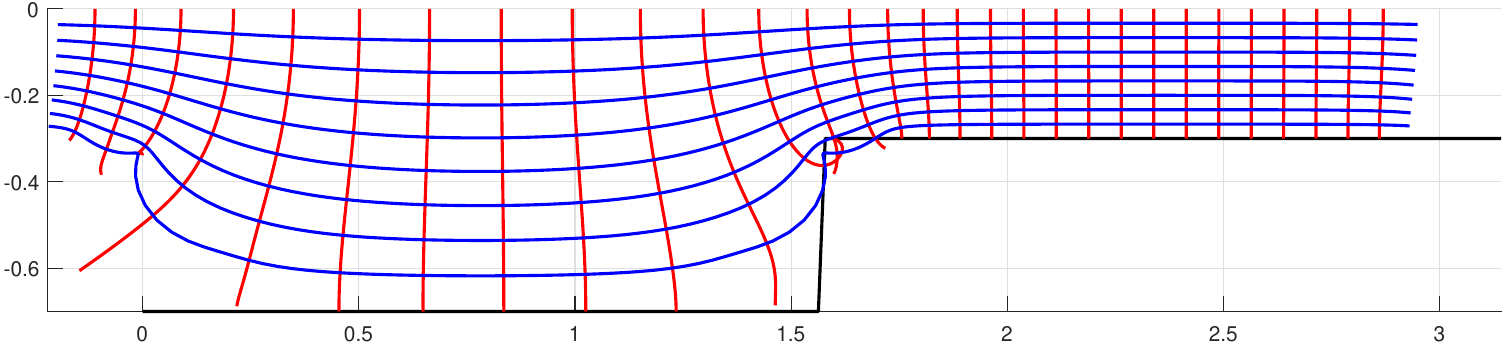}\label{fig:HHmap_examples:step}}%
	\caption{Bathymetry maps $\zmap$ from \eqref{eq:HH} with iteration procedure exemplified in \autoref{list:iterations}. Target profile is $H(\x)=.5+.1(\cos4\x+\sin6\x)$. Also shown is a diverging example of a sharp step.}%
	\label{fig:HHmap_examples}%
\end{figure}

\subsection{A piston wavemaker}
\label{sec:map:wavemaker}

Representing a piston wavemaker is remarkably straightforward with the present method---take a walled rectangular domain and shake it horizontally: 
\begin{equation}
	\zmap(\zz,t) = \zz + X(t); \qquad 
	X(t)\in\mathbb R.
	\label{eq:BM:zmap}
\end{equation}
The kinematic condition \eqref{eq:UU} then leads to
\begin{equation}
	\WW_\zz =  X_t ,
	\label{eq:BM:WW}
\end{equation}
and equation \eqref{eq:BCF} reduces to 
\begin{equation}
	\mub = 0, \qquad \mus=\left. \frac{\ppphi_\yyy}{ |\zzmap_\zzz|^2 }\right|_{\yyy=0}
	\label{eq:BM:mu}
\end{equation}
which completes  the piston wavemaker model.
The dimensions of the $\zz$-plane remain the same as those in the $\z$-plane, with $\LL = \L$ and $\hh=\h$.

\Rzero{
Since we shake the entire domain, an identical  wavemaker will generate waves at opposite phase at the opposite wall. 
These waves can be suppressed using a numerical beach, such as \eqref{eq:beach}.
Alternatively, movement of the right boundary can be avoided by replacing the translating map with one that expands and contracts: %using an expanding map rather  extending the map with some slight shrinkage and expansion:
\begin{equation}
	\zmap(\zz,t) = \zz + \rbr{1-\frac{\zz+\ii\hh}{\LL}} X(t);
	\qquad
	X(t)\in\mathbb R,
	\label{eq:BM2:zmap}
\end{equation}
where $\hh=\h$ and $\LL$ equalling the wave tank length $\L$ when the wavemaker is in its neutral position $X=0$. 
This map pulsates about the point $\z=\LL-\ii \h$ with an expansion factor 
\begin{equation}
	\alpha(t) = 1-\frac{X(t)}{\LL} = \zmap_\zz.
\end{equation}
The background velocity, satisfying  \eqref{eq:UU},  is then given by
\begin{equation}
		\WW_\zz =\rbr{1-\frac{\zz+\ii\hh}{\LL}}  \alpha X_t
		\label{eq:BM2:WW}
\end{equation}
which enters the surface boundary condition but not the bottom condition where $\Im\WW_\zz=0$:
\begin{equation}
	\mub = 0,%\zzmap_\zzz \Im\WW_\zz=	\zzmap_\zzz\frac{\h }{ L}X_t, 
	\qquad 
	\mus= \sbr{\frac{\ppphi_\yyy}{ \alpha ^2 |\zzmap_\zzz|^2 }  + 2\frac{\yy+\hh}{\alpha \LL}   \frac{\Re \zzmap_\zzz}{ |\zzmap_\zzz|^2 } X_t}_{\yyy=0}.
		\label{eq:BM2:mu}
\end{equation}
Note that $\y=0$ no longer maps to $\yy=0$ but rather to $\yy = \hh X/\alpha \LL$.
A comparison between map \eqref{eq:BM:zmap} and \eqref{eq:BM2:zmap} is provided later, in \autoref{fig:piston:compareMaps}.
}

\section{Simulation examples and validation}
\label{sec:ex}
\subsection{Steep waves on flat bathymetry (validation)}
\begin{figure}[H]
	\centering
\includegraphics[width=.5\columnwidth]{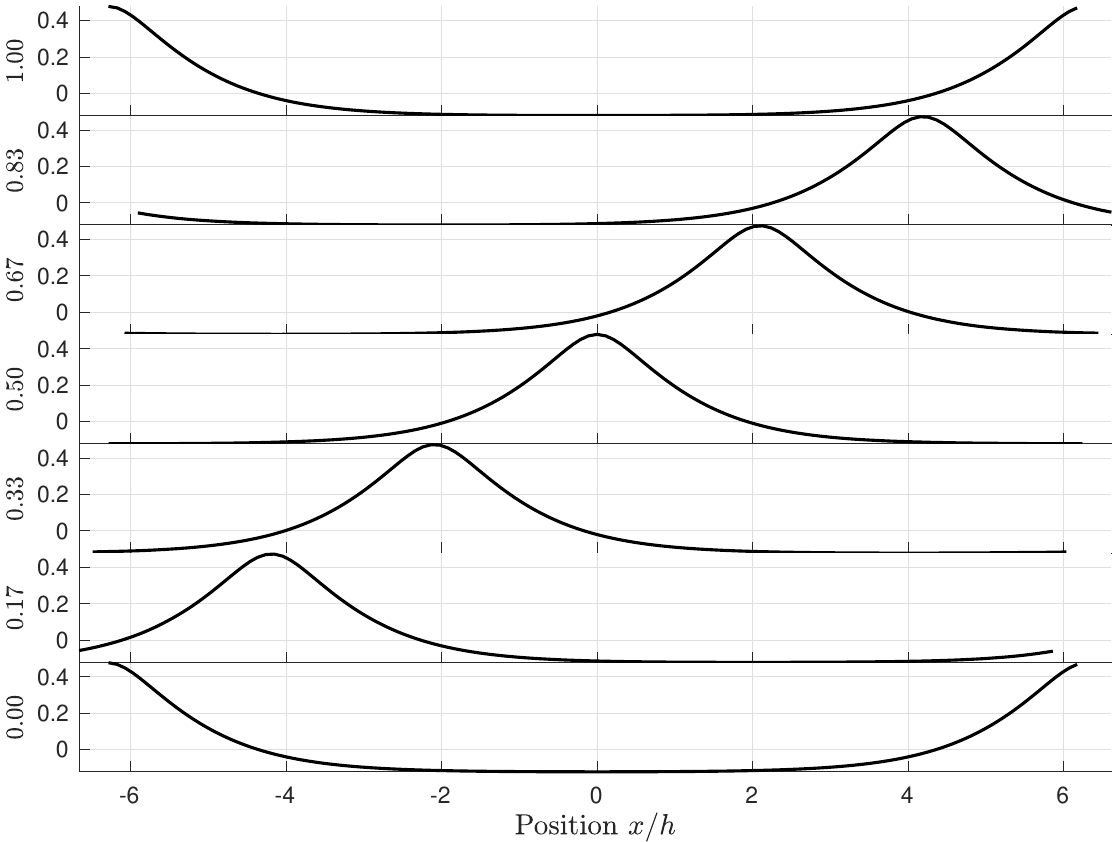}%
\includegraphics[width=.5\columnwidth]{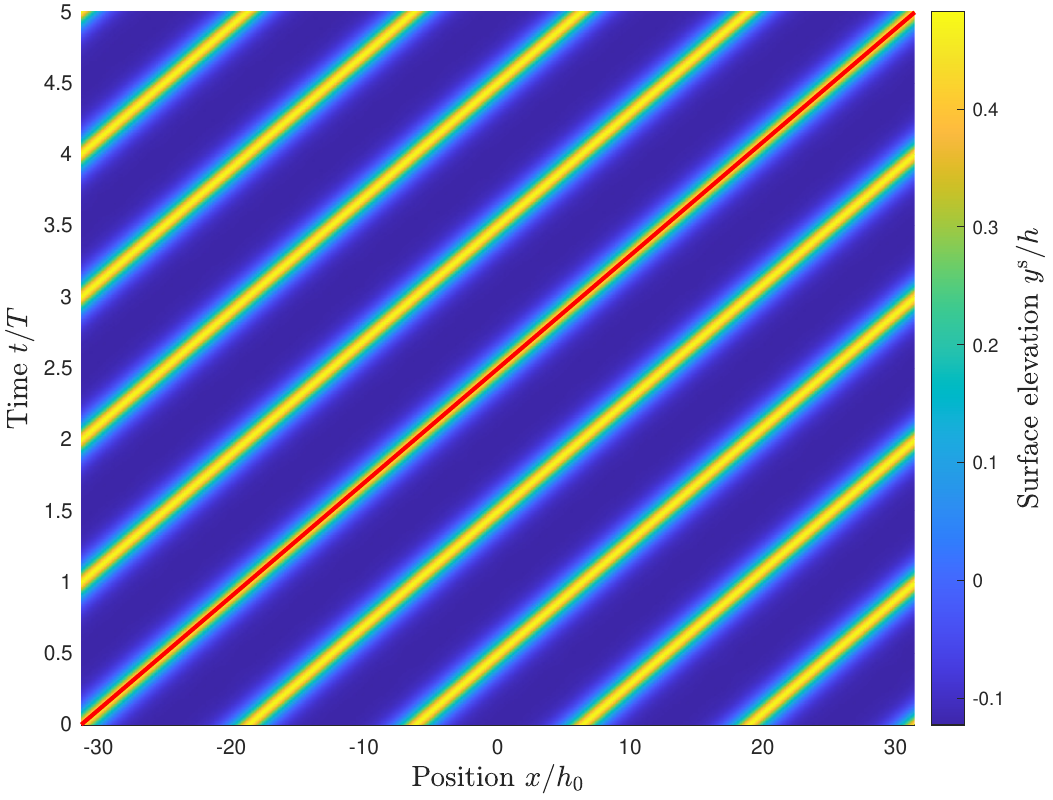}%
	\caption{Surface elevation of a shallow water stokes wave initiated from  SSGW. Red line in left panel shows the phase velocity predicted by SSGW. Steepness $kH/2=0.15$, water depth $kh=0.5$. }
	\label{fig:ssgw}
\end{figure}

As mentioned earlier, conformal mapping is particularly advantageous in shallow waters, and so shallow water Stokes waves are considered in our first benchmark.
 \autoref{fig:ssgw} shows the surface elevation of such waves in space and time, and compares their progression to the phase velocity computes with the SSGW algorithm of  \citet{clamond2018accurate}.
The initial surface elevation and potential are taken from this reference, and five wave crests are simulated over five periods in a periodic domain, requiring about two CPU seconds. 
The waves maintain their shape and propagate with expected celerity (red line).

\begin{figure}[H]
	\centering
	\subfloat[Wave steepness $kH/2 = 0.05$]{\includegraphics[width=.5\columnwidth]{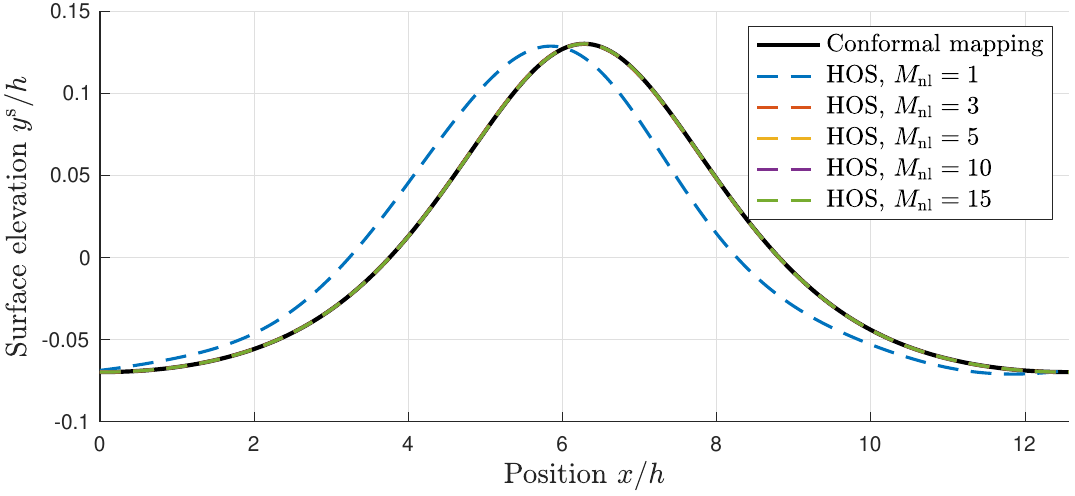}}%
	\subfloat[Wave steepness $kH/2 = 0.10$]{\includegraphics[width=.5\columnwidth]{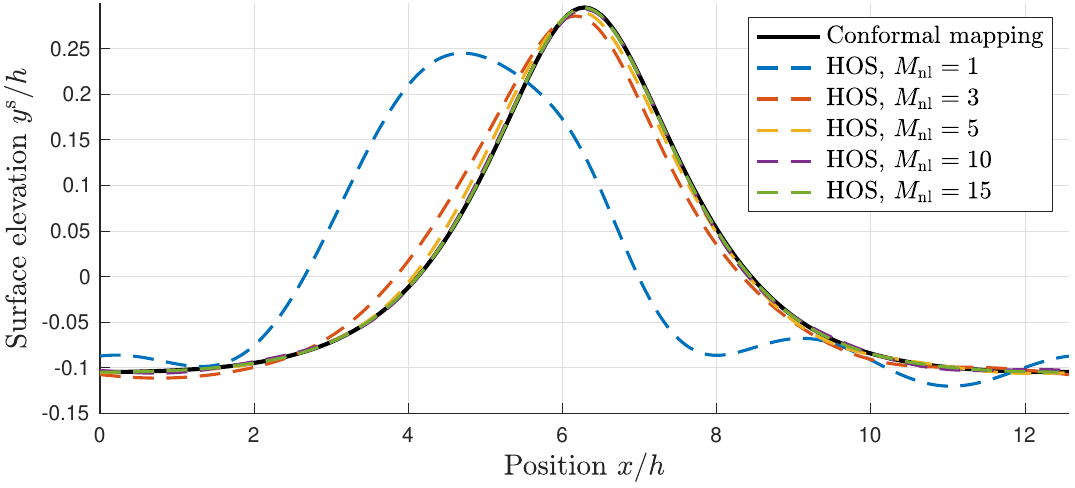}}\\
	\subfloat[Wave steepness $kH/2 = 0.15$]{\includegraphics[width=.7\columnwidth]{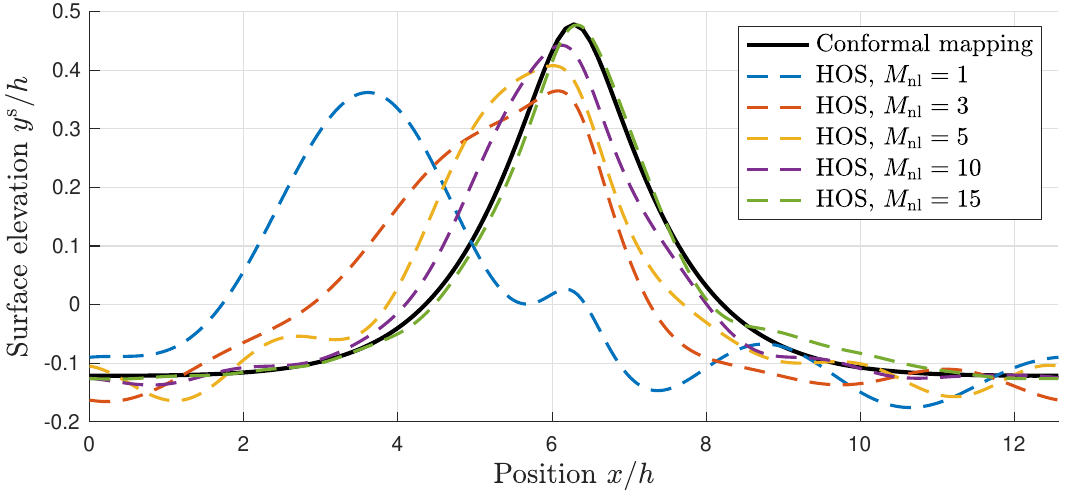}}
	\caption{Comparison to simulations with the HOS method at various orders of nonlinearity $M\_{nl}$; surface elevation after simulating five wavelengths for five periods, cf.\ \autoref{fig:ssgw}.  Initial state  generated with SSGW. Water depth $kh = 0.5$.}
	\label{fig:HOS}
\end{figure}

HOS methods are powerful tools for intermediate water depths and three-dimensional wave fields. 
Accuracy and efficiency are however closely tied to the required order of HOS nonlinearity, $M\_{nl}$, 
the number of FFT operation per time step being $4+ M\_{nl}(M\_{nl}+3)/2$, plus an additional four for filtering.
Repeating the previous simulation (five waves, five periods, $kh=0.5$) for varying wave steepness produces the surface profiles shown in \autoref{fig:HOS}. 
A nonlinearity order of three accurately captures the moderately steep wave  $kH/2=0.05$, while orders between five and ten are needed for $kH/2=0.10$. 
The steepest wave, $kH/2=0.15$, converges almost but not quite with 15 orders and diverges beyond that. 
Various HOS method versions exist today, and alternative initial conditions---such as gradually increasing model nonlinearity over time---may yield better results. 
Nevertheless, this example highlights the inherent advantage of the conformal mapping method over the HOS model in shallow waters.
Conformal mapping requires seven FFT transformations per times step, plus an additional four for the stabilisation \eqref{eq:damping}.

\begin{figure}[H]
	\centering
	\includegraphics[width=.7\columnwidth]{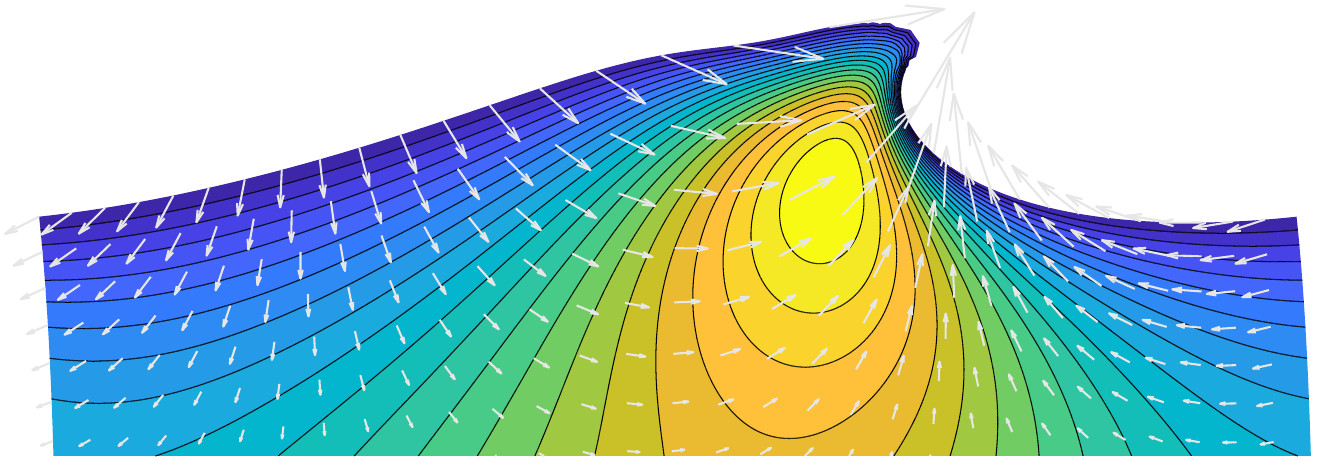}%
	\caption{Breaking wave case corresponding to figure 2 in \citet{chalikov2005modeling}. 
		Contours show deviation from the generalised hydrostatic pressure $p+g\yyy$.
		(Image tapered from below.)}
	\label{fig:ChalikovWave}
\end{figure}

While considering flat bathymetries, it is natural to reproduce one of the more striking cases presented by \citet{chalikov2005modeling}.
\autoref{fig:ChalikovWave} captures a snapshot from a simulation whose initial conditions were that of an extremely steep linear wave crest ($ka=0.5$) within a periodic domain.
Shortly after initiation, the wave breaks, and the snapshot is taken just before the simulation crashes.
Our essentially parametric surface description permits a slight folding of the wave crest.
However, the dynamics of the plunging wave front cannot be captured within the framework of potential theory. 
The contours in \autoref{fig:ChalikovWave} display, for comparison to reference, the `deviation from the generalised hydrostatic pressure', whose difference  from the dynamic pressure is $g\yyy-g\y$.
Consistent with our reference, we observe the same pressure `bubble' underneath the wave front. 
The water depth of this case is  $kh=\pi$, with stabilisation parameters $\kd = 0.5\,\kMax$ and $\rDamping = 0.25$, using a padding factor of 4.5.

\subsection{Piston wavemaker (map of \autoref{sec:map:wavemaker})}
\label{sec:ex:piston}
The simple mapping described in \autoref{sec:map:wavemaker} enables arbitrary wavemaker motions, typically composed of random Fourier components distributed according to an energy spectrum.
For testing purposes, we  consider a single-harmonic wavemaker motion,
 $X_0(t) = \hat X_0 \sin \omega_0 t$, with an added ramp $R(t)$ to softens the initial motion:
 $X(t)=X_0(t)R(t)$, $X_t=\FFInv\sbr{\ii\omega\FF(X_0R)}$, and so forth.

\begin{figure}[H]
	\centering
	\subfloat[{Fluid particle position; $\Re [\z\^p(t)] = X(t)$.}]{\includegraphics[width=.5\columnwidth]{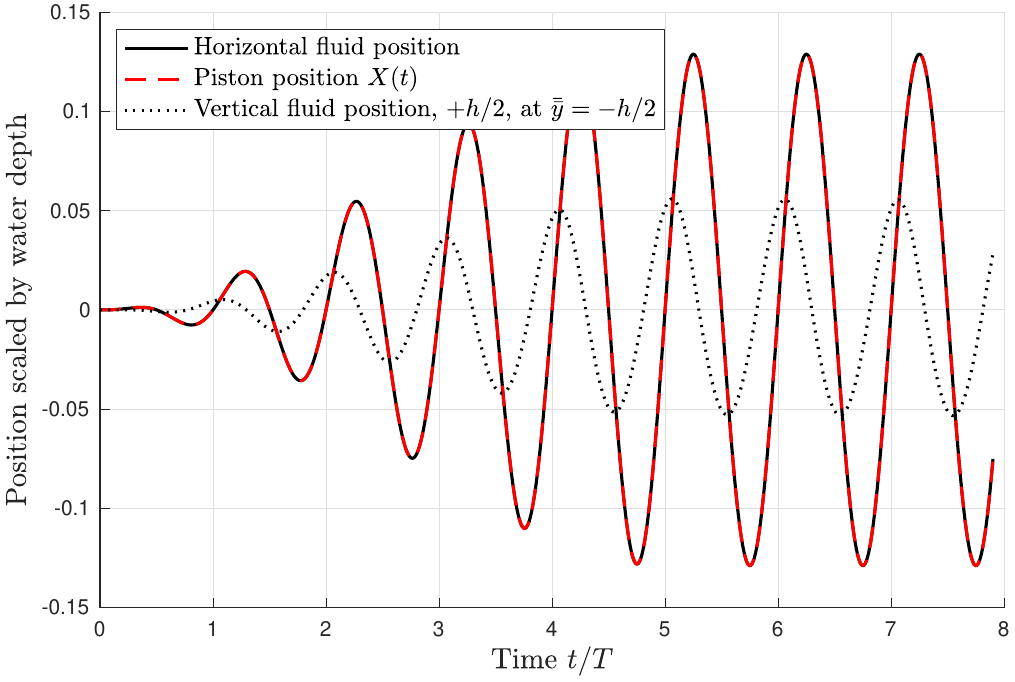}}%
	\subfloat[{Fluid velocity; $\phi_\x[X(t),\y,t]=X_t(t)$.}]{\includegraphics[width=.5\columnwidth]{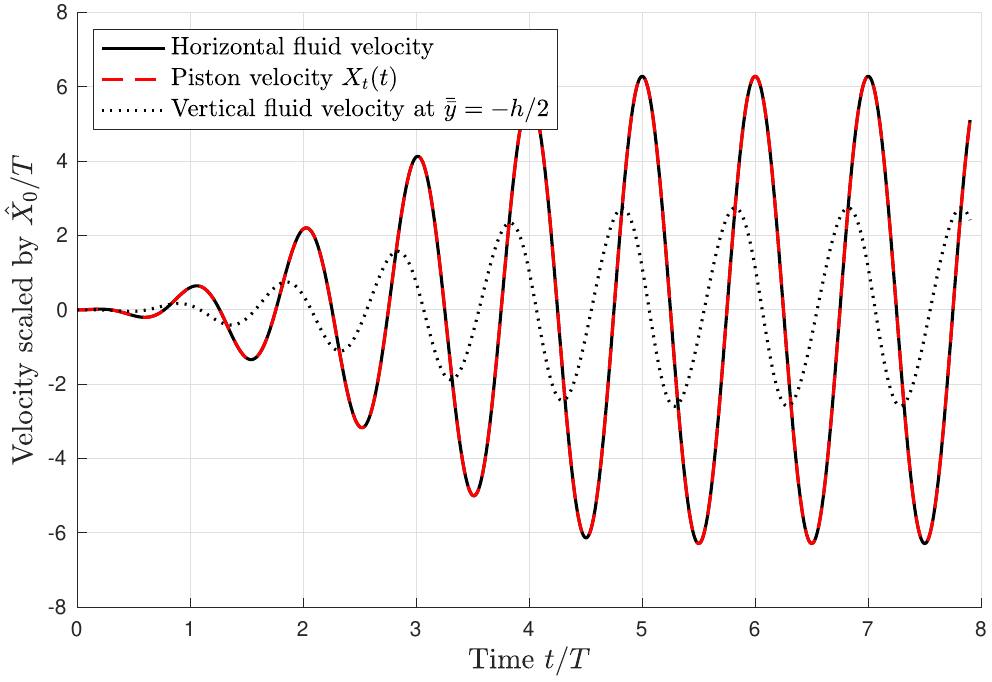}}%
	\caption{Validation of the piston boundary condition.}
	\label{fig:piston:BC}
\end{figure}

First, we verify the boundary conditions.
As shown in  \autoref{fig:piston:BC} , the condition $\phi_\x[X(t),\y,t]=X_t(t)$ is exactly satisfied per map design. 
Vertical velocities in the mid water depth range are also shown in the figure. 

\begin{figure}[H]
	\centering
	\includegraphics[width=\columnwidth]{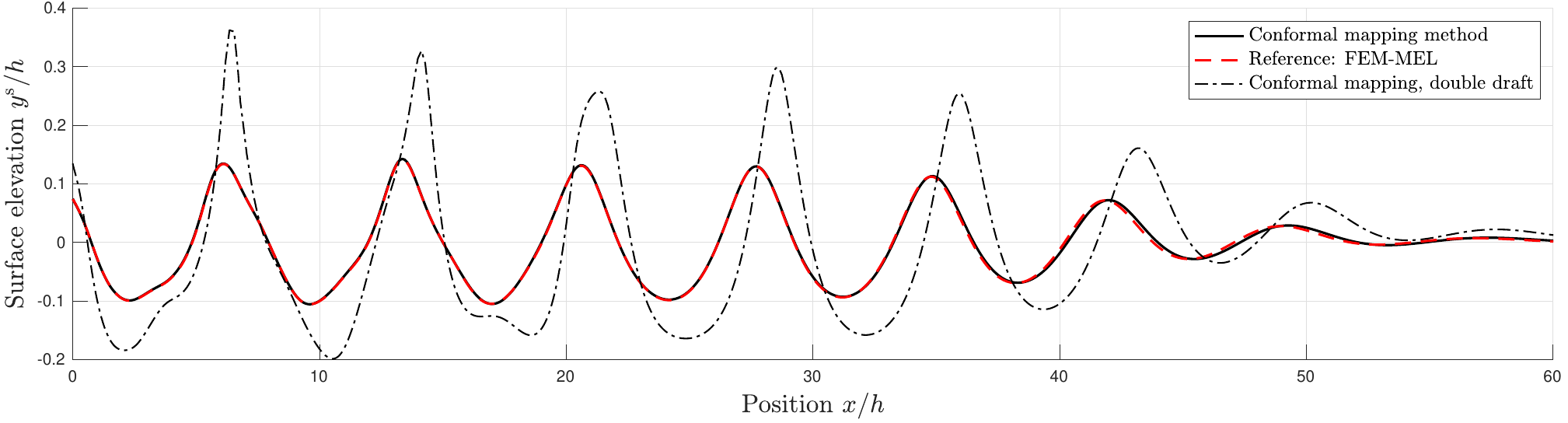}%
	\caption{Validation of simulated surface elevation against. The period is $T=7.93\sqrt{h/g}$.}
	\label{fig:piston:eta}
\end{figure}

The surface elevation for this case is the depicted in \autoref{fig:piston:eta}, along with a validation comparison against one of SINTEF Ocean's in-house codes.
This in-house model is a sigma-grid-based FEM  \citet{wu1995_FEM}, incorporating Lagrangian particle support. It has, in turn, been benchmarked against other FDM codes.
The agreement in \autoref{fig:piston:eta} is excellent but for a slight difference in phase velocity as seen at the front end of the wave train. 
This  discrepancy originates in fact from grid stretching errors in the reference model---a well documented limitation of sigma-grid transformations. 
Grid stretching error can be minimised through clever meshing \citep{pakozdi2022_gridStretching} given prior knowledge of phase velocity.

The wavemaker example highlights the core challenge of wave generation:
 a single-harmonic wavemaker motion  inevitably produces spurious super-harmonic waves alongside the primary wave. 
 These propagate at slower group velocities and are therefore observed at the back end of the wave train, leaving the front end unaffected.
 These will, as seen in the figure,  propagate at slower group velocities. 
 Wave contamination is therefore observed at the back-end of the wave train, leaving the front-end clean.
To further emphasise this effect, a second case with double the piston stroke is included in \autoref{fig:piston:eta}, shown in a dot-dashed line.
The simulations run in a few seconds, whereas the FEM/FDM codes require several orders of magnitude more computation time.

\begin{figure}[H]
	\centering
	\subfloat[Translatory map \eqref{eq:BM:zmap}.]{\includegraphics[width=.5\columnwidth]{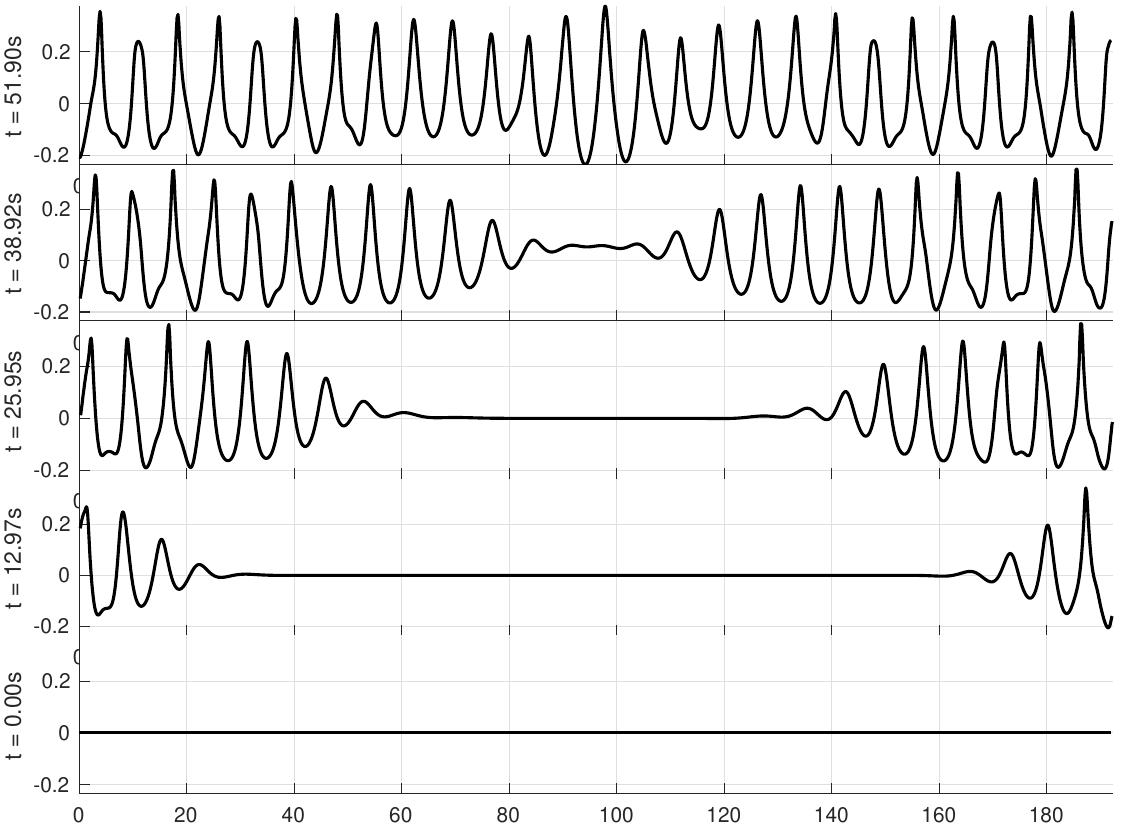}}%
	\subfloat[Contracting map \eqref{eq:BM2:zmap}.]{\includegraphics[width=.5\columnwidth]{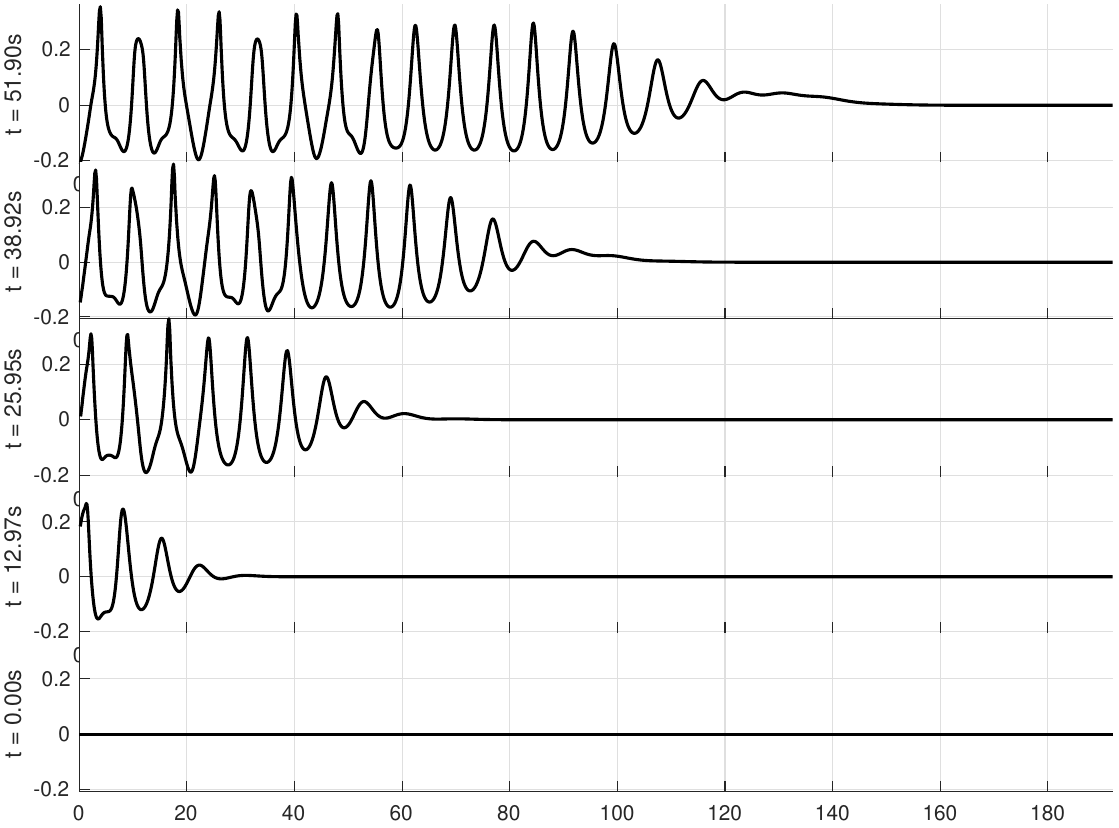}}%
	\caption{
		\Rzero{Full domain simulated in above figures, comparing the two wavemaker maps in \autoref{sec:map:wavemaker}. Case with double piston draft. No numerical beach included.}}
	\label{fig:piston:compareMaps}
\end{figure}

\Rzero{
Both wavemaker mappings introduced in   \autoref{sec:map:wavemaker} yield the results shown, which display only the region near the wavemaker.
The difference between the mappings become evident when zooming out, as in \autoref{fig:piston:compareMaps}, to include the beach-side section of the wave tank. 
With mapping  \eqref{eq:BM:zmap}, waves of  opposite in phase and direction are being generated at the beach end of the domain. 
No beach-side wave generation occurs with mapping \eqref{eq:BM2:zmap}.
}

\subsection{A tsunami (map of \autoref{sec:map:SC})}
\newcommand{\Tf}{T\_f}
\newcommand{\Dh}{\Delta h}

We now turn to testing transient bathymetries selecting the case of a tsunami triggered by the uplift of a tectonic shelf, modelled using the analytically integrated map  \eqref{eq:map_logstrip}.
The initially flat shelf rises during a transition period $\Tf$ until reaching a final height $\Dh$. 
For ease of validation, the rise velocity is made constant. % , equalling $\Dh/\Tf$.

Map  \eqref{eq:map_logstrip}  is applied for this test, with time dependency introduced by defining  $h_1(t) = h_2 - \Dh \min(t/\Tf,1)$. 
The time derivative is computed using discrete differentiation $\zmap_t=[\zmap(\zz,t+\epsilon/2)-\zmap(\zz,t+\epsilon/2)]/\epsilon$.
The mirroring approach \eqref{eq:mirrorWall} is applied to  the lateral boundaries, but without an added background potential. 
This  enforces zero horizontal velocity along lateral boundaries, even as the left boundary shifts in time. 
The boundary then remains stable and does not interfere with the simulation, provided wave never reach the domain edges.

\begin{figure}[H]
	\centering
	\subfloat[Vertical velocity component.]{\includegraphics[width=.5\columnwidth]{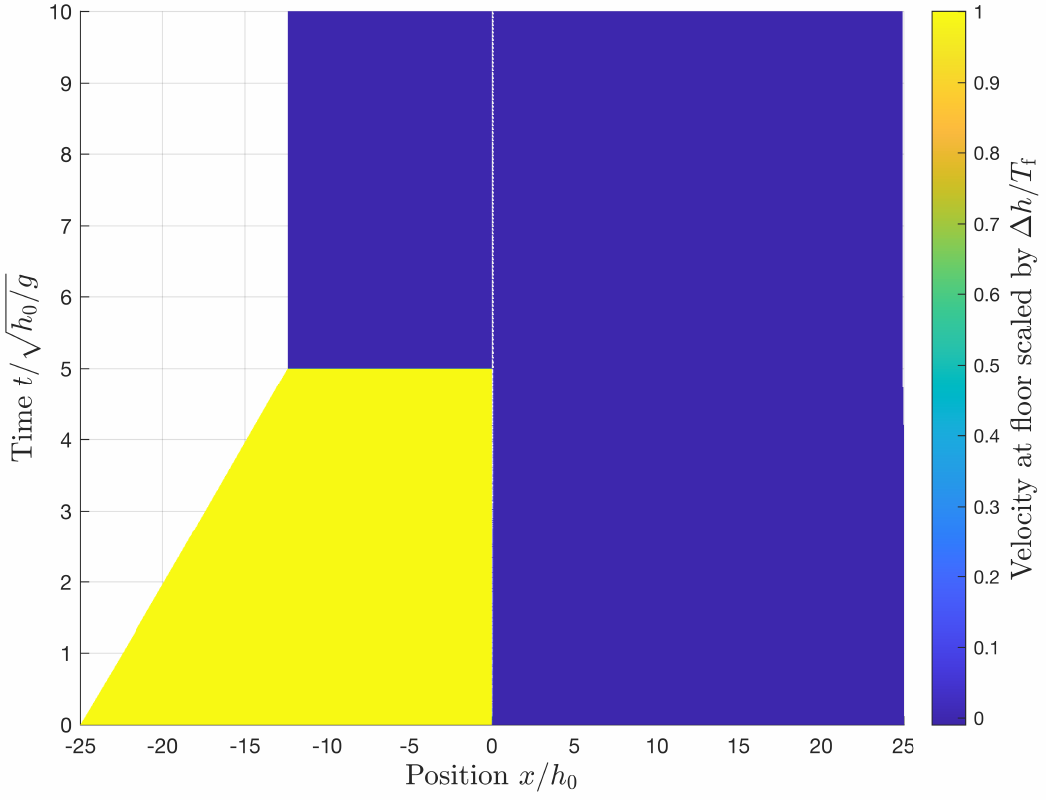}}%
	\subfloat[Horizontal velocity component.]{\includegraphics[width=.5\columnwidth]{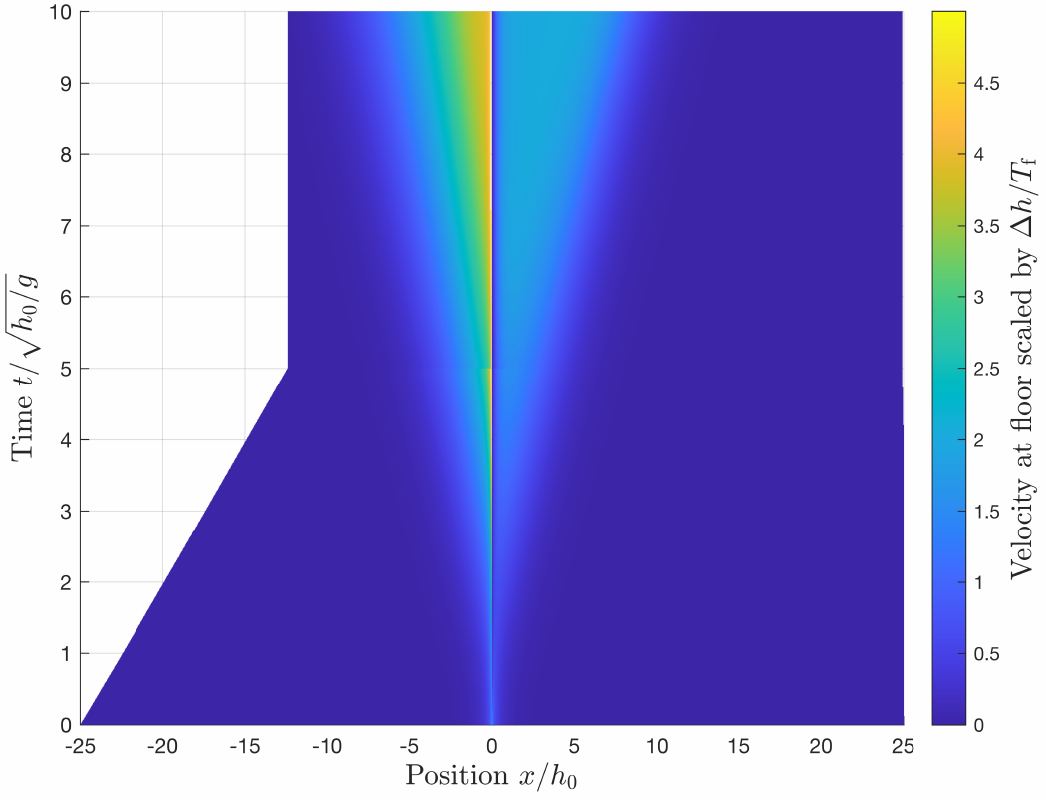}}%
	\caption{Velocity at bed, scaled by the prescribed shelf rise speed $\Dh/\Tf$. $\Dh = 0.50 \, h_0$, $\Tf = 10.0 \sqrt{h_0/g}$.}
	\label{fig:tsunami:validation}
\end{figure}

We again start by verifying the  kinematic boundary conditions, this time at the bed. 
\autoref{fig:tsunami:validation} presents the velocity field $\w_\z^*$ for a representative case. 
As required, the vertical velocity remains zero in the right half of the domain, while in the left half, it equals the shelf rise velocity $\Dh/\Tf$ during the transition period $\Tf$, returning to zero afterwards.
The left boundary also shifts rightward as the shelf rises, due to coordinate compression in this region.
This reduces the length of the domain but causes no other issues.
\Rzero{
If preferred, the shift can be greatly reduced by choosing a mapping function which, instead of compressing the left domain, rises it.
A suitable alternative is the simpler Schwartz--Christoffle step transformation \eqref{eq:SCPrimitive}.
}

\begin{figure}[H]
	\centering
	\subfloat[Light case; $\Dh = 0.25 \,h_0$, $\Tf = 1.0 \sqrt{h_0/g}$.]{%
		\includegraphics[width=.5\columnwidth]{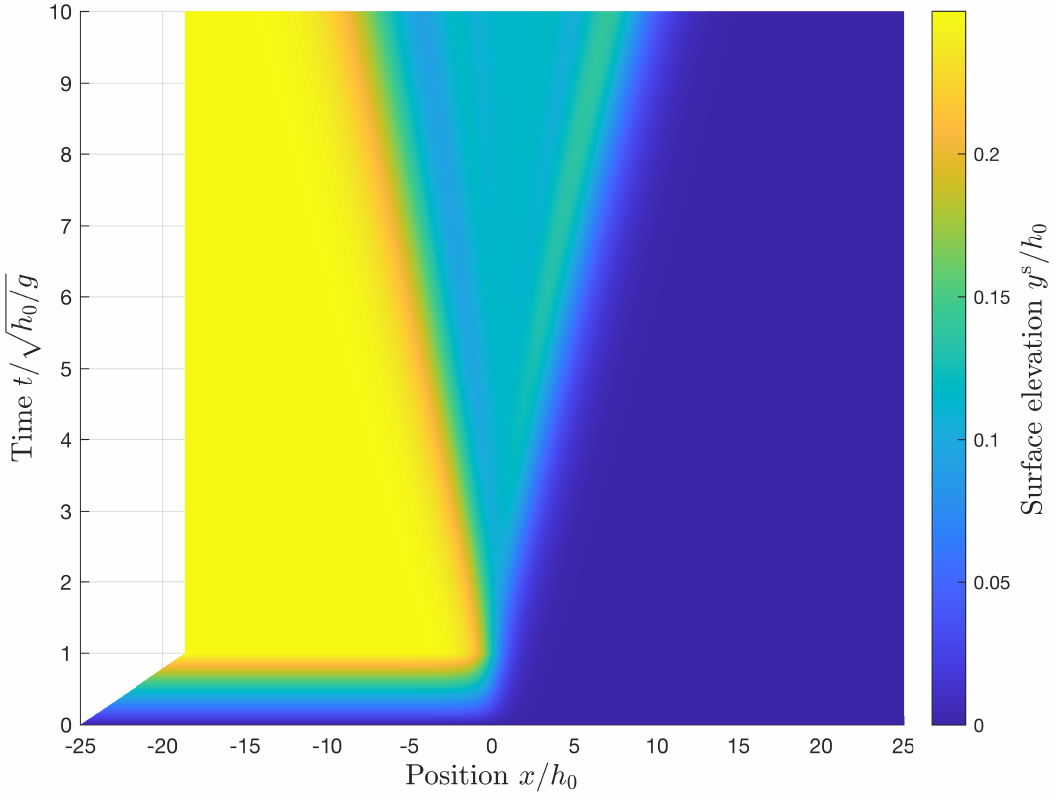}%
		\includegraphics[width=.5\columnwidth]{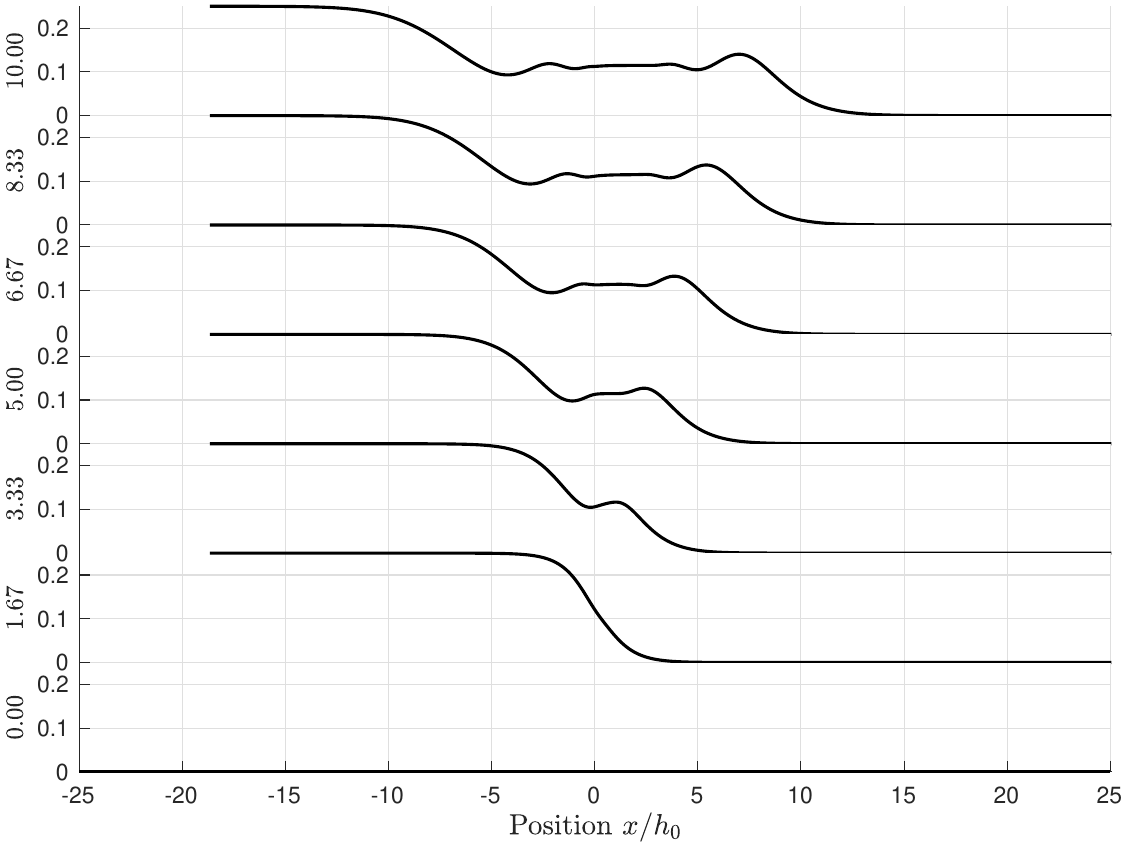}\label{fig:tsunami:eta:light}}%
	\\
			\subfloat[Slow case; $\Dh = 0.50\, h_0$, $\Tf = 10.0 \sqrt{h_0/g}$.]{%
		\includegraphics[width=.5\columnwidth]{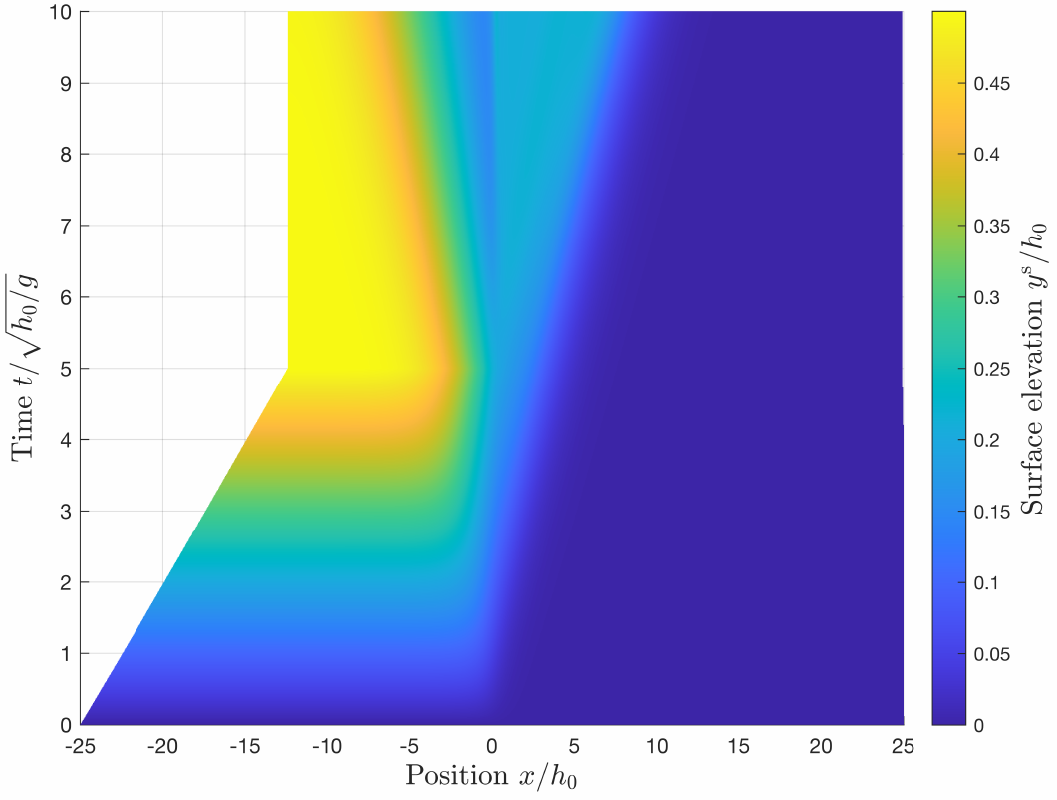}%
		\includegraphics[width=.5\columnwidth]{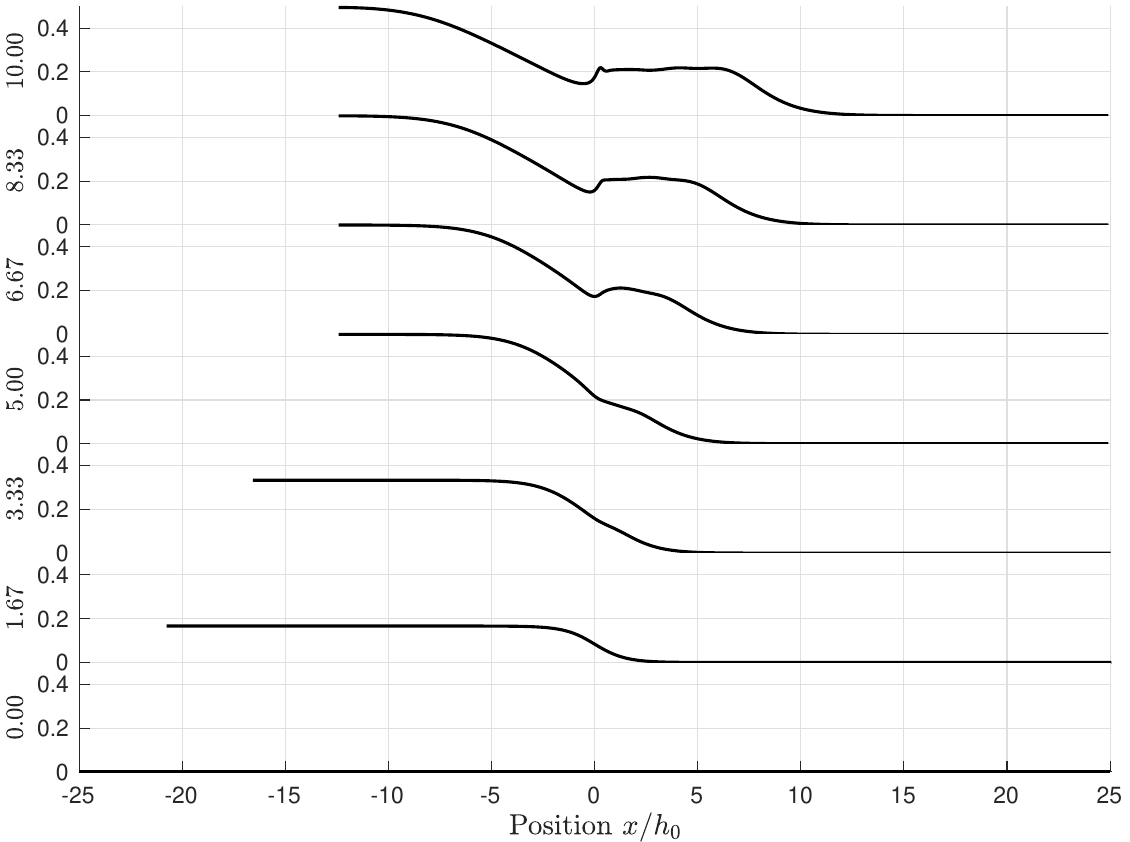}\label{fig:tsunami:eta:slow}}%
	\\
	\subfloat[Abrupt case; $\Dh = 0.50\, h_0$, $\Tf = 0.5 \sqrt{h_0/g}$.]{%
		\includegraphics[width=.5\columnwidth]{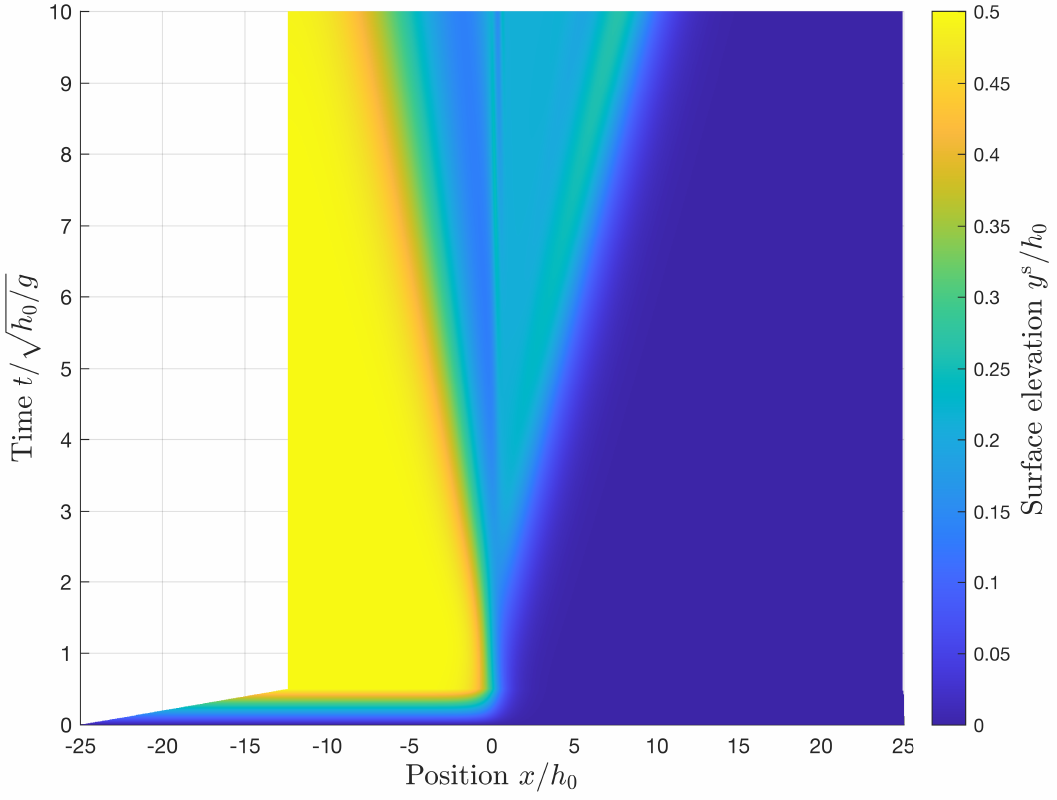}%
		\includegraphics[width=.5\columnwidth]{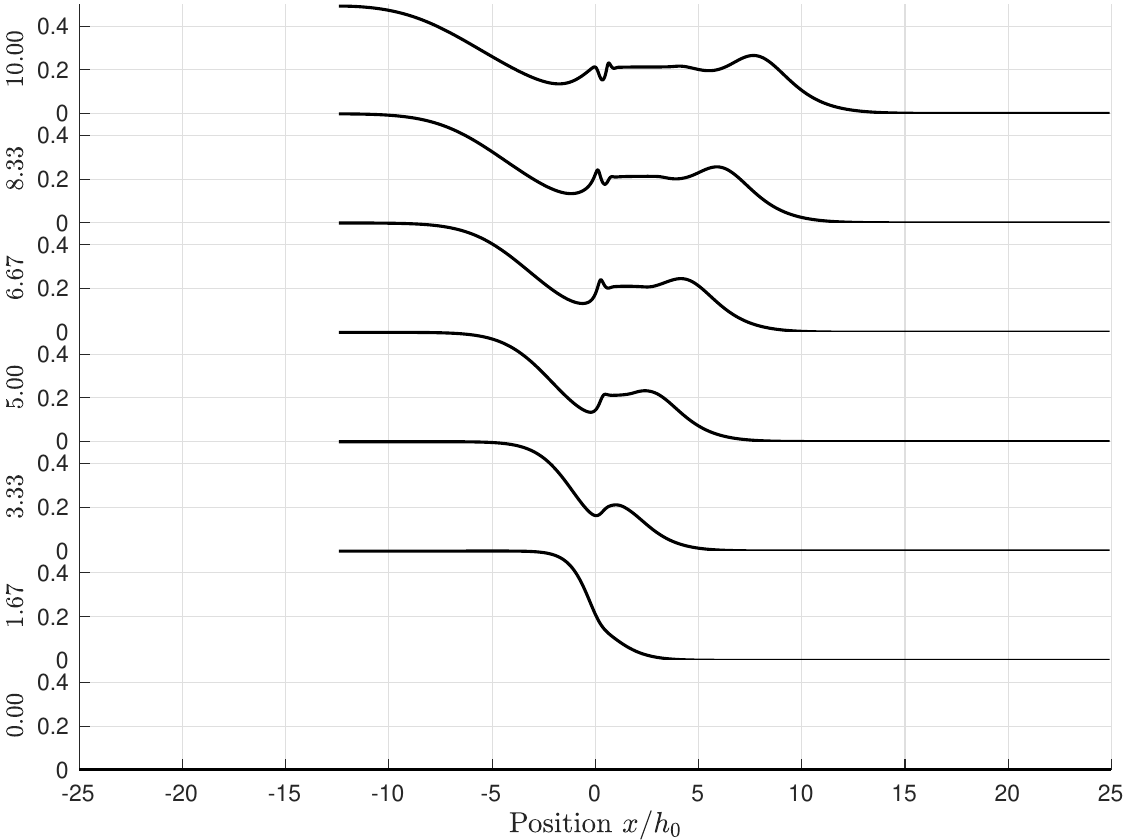}\label{fig:tsunami:eta:hard}}%
	\caption{Tsunami example; surface elevation caused by a bottom shelf rising at fixed velocity $\Dh/\Tf$ during a period $\Tf$.}
	\label{fig:tsunami:eta}
\end{figure}

\autoref{fig:tsunami:eta} illustrates surface elevations for varying values of $\Dh$ and $\Tf$.
The solutions exhibit a far-field plateau that follows the bathymetric rise.
In between, an intermediate plateau fans out, connecting to the far fields through a rarefaction wave and a softened shock front.
Waves generated by transient dynamics occupy this middle plateau. These form secondary wave fronts when the shelf rises abruptly. 

\begin{figure}[H]
	\centering
	\subfloat[$t = 0.5$ and $1.0\,\Tf$.]{\parbox{.5\columnwidth}{
	\includegraphics[width=.5\columnwidth]{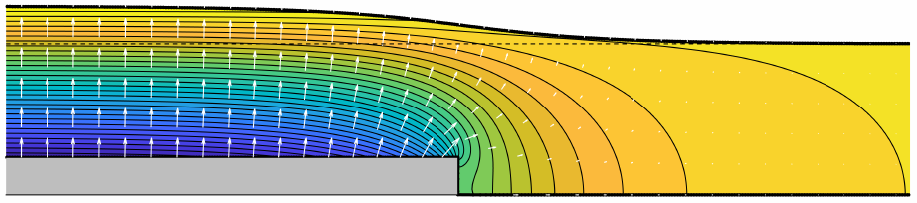}\\
	\includegraphics[width=.5\columnwidth]{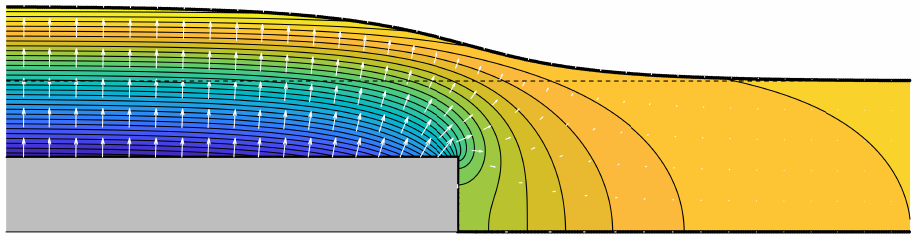}}}%
	\subfloat[$t = 1.5$ and $2.0\,\Tf$.]{\parbox{.5\columnwidth}{
	\includegraphics[width=.5\columnwidth]{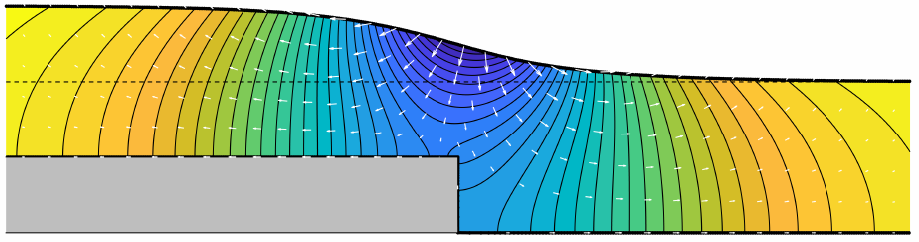}\\
	\includegraphics[width=.5\columnwidth]{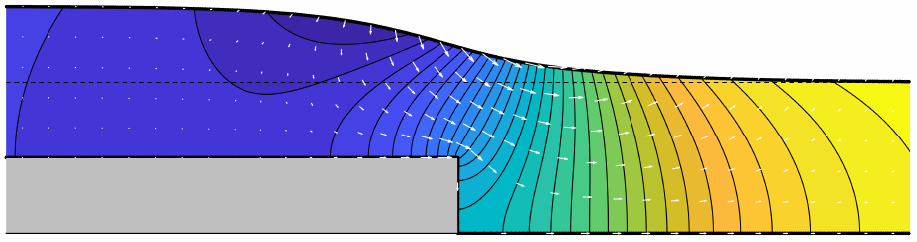}}}\\
	\subfloat[$t=20\,\Tf$ (end time).]{
	\includegraphics[width=.5\columnwidth]{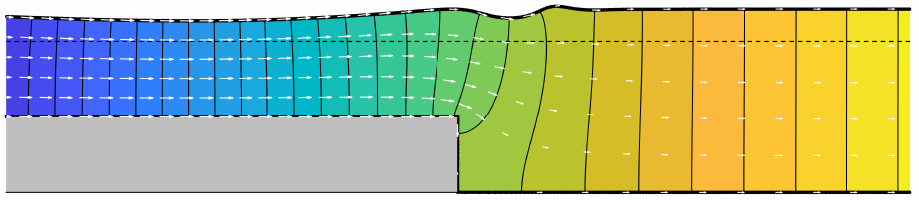}}%
	\caption{Velocity and potential field reconstructed from surface variables. Abrupt rise case shown in \autoref{fig:tsunami:eta:hard}}
	\label{fig:tsunami:field}
\end{figure}

Snapshots of the interior velocity field are shown in \autoref{fig:tsunami:field}  for the most abrupt rise.
Field recreation uses  \eqref{eq:www} and \eqref{eq:zzzMap}, requiring only  the prescribed mapping function and surface variables $\etazzOfxxx$ and $\ppphi\^s$. 
As with previous cases, computation time remains within the scale of seconds, depending on resolution and the intensity of transients. 
The dynamic ODE solver adopts fine time stepping during the depth transition and then quickly speeds up once the bed reaches its stationary state.

\subsection{Wave reflection across a depth transition ramp (map of \autoref{sec:map:SCnum})}
\begin{figure}[H]
	\centering
	\subfloat[$30\degree$ beak angle.]{\includegraphics[height=.3\textwidth]{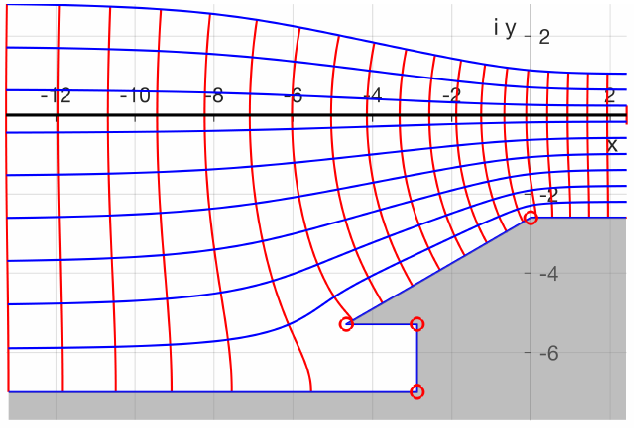}}%
	\hfill
	\subfloat[$15\degree$ beak angle.]{\includegraphics[height=.3\textwidth]{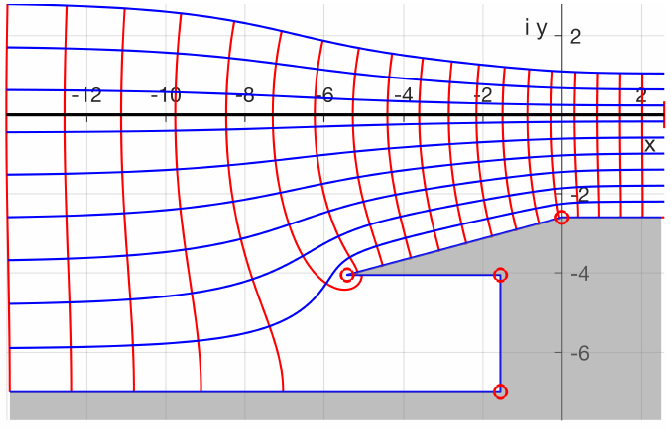}}%
	\caption{Beak geometry maps with a depth transition from $7.0$ to $2.6$ meters.}
	\label{fig:beak:geometry}
\end{figure}

\begin{figure}[H]
	\centering
		\subfloat[Surface elevation as packet passes across depth transition.]{\includegraphics[width=.5\textwidth]{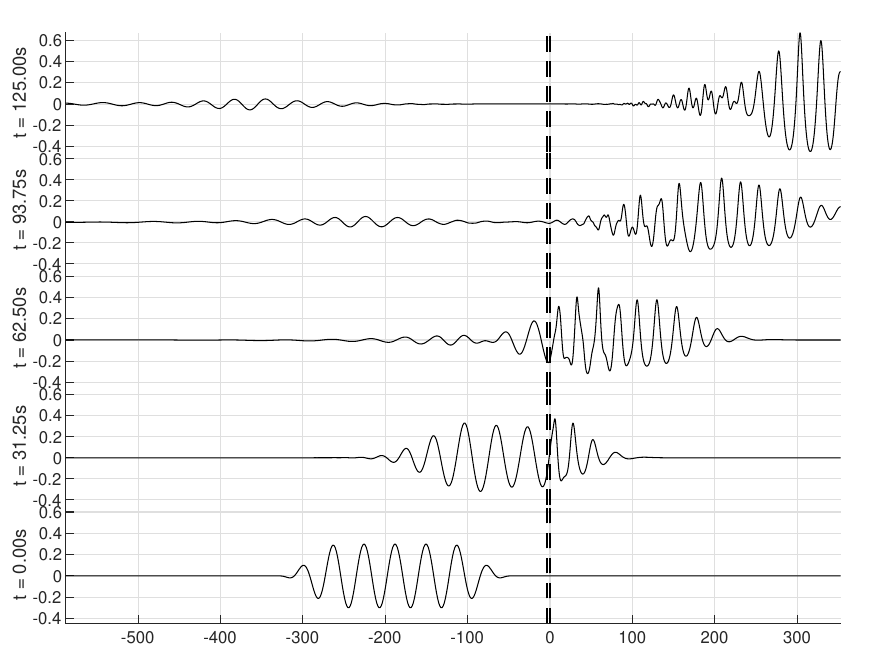}\label{fig:beak:eta}}%
		\subfloat[Equation \eqref{eq:Cr} gauging the wave energy contained in the deep side of the domain.]{\includegraphics[width=.5\textwidth]{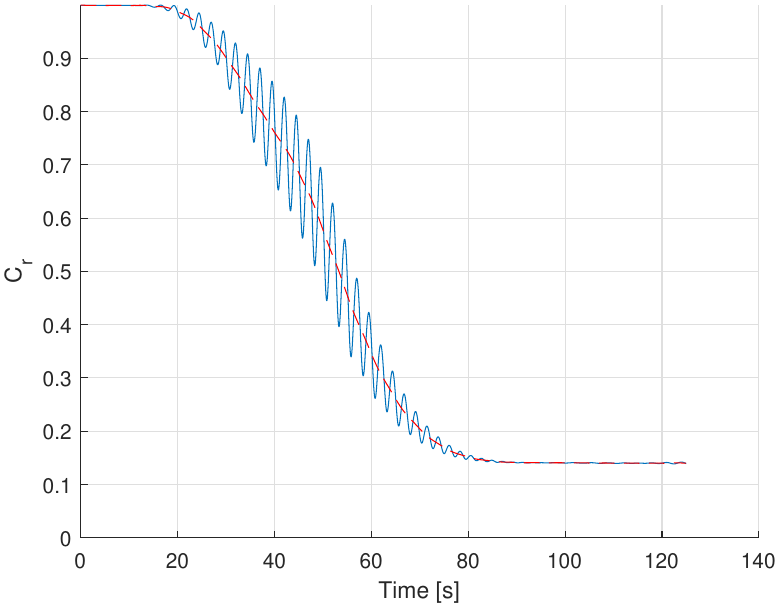}\label{fig:beak:Crt}}%
	\caption{Illustration of analysis for example period $5.0$ seconds, 45\textdegree{} beak angle, depth transition from $12.0$ to $2.6$ meters.}
	\label{fig:beak:t}
\end{figure}

\begin{figure}[H]
	\centering
		\subfloat[Wide range of angles.]{\includegraphics[width=.5\textwidth]{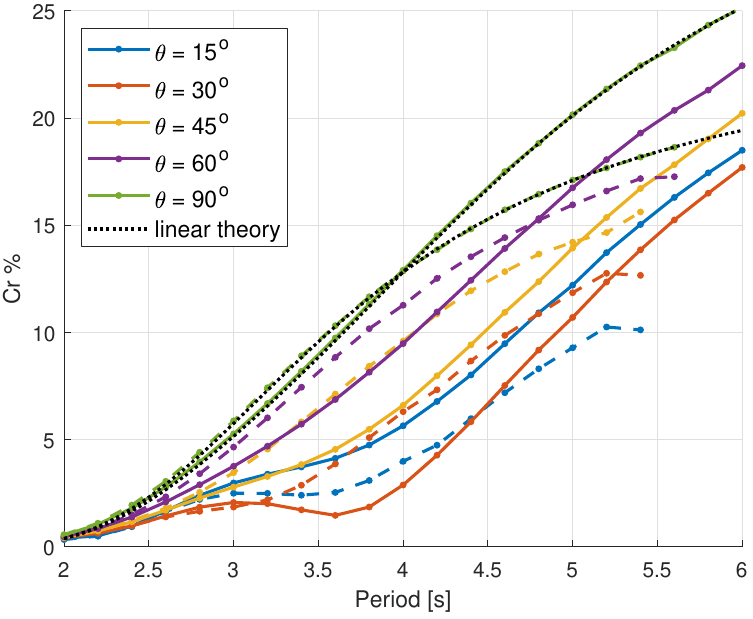}\label{fig:beak:Cr:wide}}%
		\subfloat[Refined range of angles.]{\includegraphics[width=.5\textwidth]{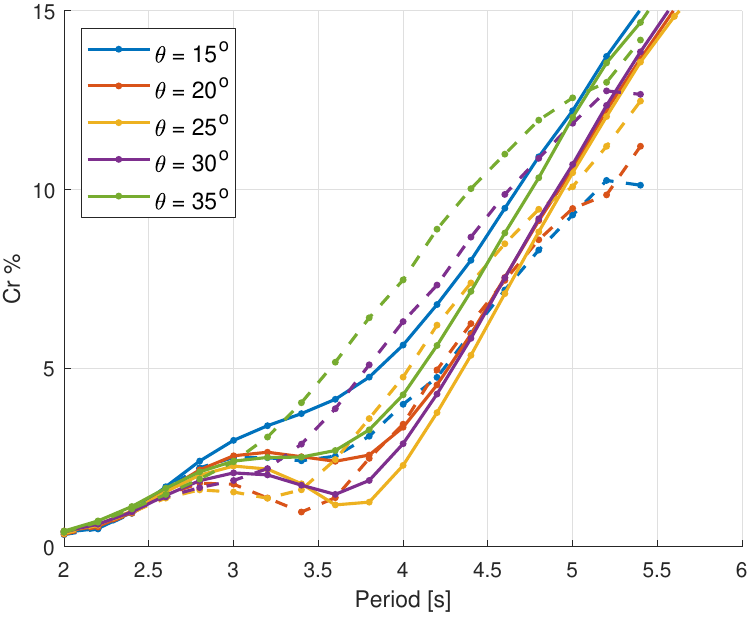}}%
	\caption{Reflection curves as function of wave period. Transition from two depths, 12.0 meters (solid lines) and 7.0 meters (dashed lines), are shown, shallow water depth being 2.6 meters. A reference estimate from linear theory for $\theta=90\degree$ is also included. Beak angles are displayed in the legend. }
	\label{fig:beak:Cr}
\end{figure}

As a final example, a small study made at SINTEF Ocean in Norway is briefly reproduced.
The study was part of the design process for the Norwegian Ocean Technology Centre, which is currently under construction.
It investigated whether installing a beak-shaped extension at the edge of an abrupt depth transition could reduce wave reflections; two geometry examples are shown in \autoref{fig:beak:geometry}.
Simulations were performed using the prototype model  \eqref{eq:prototype}, varying  beak angles and deep-side water depths while keeping the beak length and shallow-side depth fixed. 
Short wave packets were initialised and allowed to propagate across the depth transition. 
Having passed the transition, all the wave energy that remains on the deep-water side is assumed to be reflected energy.
Integrating this, reflection coefficient  
\begin{equation}
	C\_r(t) = \cbr{ \int_{-\infty}^0\![\y\S(x,t)]^2\,\dd x \bigg/ \int_{-\infty}^0\![\etaz(x,0)]^2\,\dd x  }^{1/2}
	\label{eq:Cr}
\end{equation}
is computed.
\autoref{fig:beak:t} illustrates the process and the evolution of $C\_r(t)$ over time as the packet transitions. 
A final  reflection value was determined as the filtered minimum of $C\_r(t)$.
The process was easily automated, enabling rapid generation of the reflection chart in \autoref{fig:beak:Cr}.
Each simulation  required about 30 to 60 CPU seconds, with a full reflection curve computed in 10 to 20 minutes.
The study confirmed that the beak extension reduces reflection, with an optimal beak angle that depends on the water depth---approximately 25\textdegree{} for a 12 meter depth and 20\textdegree{} for 7 meters.

A final  benchmark is included in \autoref{fig:beak:Cr:wide};
dotted black lines show the predictions from linear mode-matching theory \citep[e.g.,][]{li_2021_step1}, which applies to straight step transitions $\theta=90\degree$.
The reference model shows remarkable agreement, especially  considering the simplicity of estimate \eqref{eq:Cr}. 
To avoid wave breaking, the initial wave steepness is a moderate $kH/2=0.05$ in all cases. Waves undergo significant steepening after the depth transition, increasingly so for larger periods.
The release and reflection of second-order bound harmonics are clearly seen in \autoref{fig:beak:eta} as secondary wave packets, while
higher-order free wave packets become visible with increasing steepness.
Better simulation performance and steeper waves can be obtained with the full conformal mapping model.


\section{Closing remarks}%{Summary and future development}
\label{sec:summary}
A water wave simulation method based on conformal mapping has been presented, along with a range of practical examples. 
The method demonstrates high accuracy in all cases and never requires more than a minute of CPU time. 
At its core, the model simulation routine follows a compact series of explicit steps. 
Initialisation involves  additional routines for setting  initial conditions and mapping of the domain geometry, but these introduce negligible computational overhead. 
Efficiency is strongly dependent on resolution, wave steepness, the intensity of transients, numerical stabilisation and the extent of anti-aliasing, and so a more precise measure of efficiency has therefore not been pursued here.

Development of a wave breaking model within the conformal mapping framework is recommended.
While dissipation model \eqref{eq:damping} effectively suppresses high-wavenumber energy buildup, it is ad hoc e and should not be relied upon to represent the propagation and dissipation process of a breaking wave front.
Instead, wave breaking models should be incorporated into the conformally mapped model, drawing from established studies on wave-breaking dynamics. 
\citet{seiffert2017_waveBreakingOnsetInHOS,seiffert2018_waveBreakingHOS} have already successfully integrated such a model into the HOS scheme, using the breaking onset criterion of \citet{barthelemy2018_waveBreakingModel} and the eddy dissipation models of \citet{tian2010_waveBreakingEddieViscosity}.
Conceptually, 
the detailed information about crest evolution curvature that is available in the conformal representation could be advantageously utilized in this respect.

Conformal mapping has great potential for wavemaker modelling
and, to the author's knowledge, this represents the first wavemaker model of its kind.
\Rzero{In addition to the piston-type wavemaker models presented here, 
	the author has recently developed a a paddle-type wavemaker models, which has been successfully validated.   
	Due to its scope and potential impact, its details are reserved for a forthcoming paper that builds upon the present study.
	The paddle wavemaker completes the conformal mapping model as a fully functional numerical wave tank, enabling integration with most wave generation systems.
	}

\section*{Acknowledgements}
The authors would like to acknowledge support from Statsbygg, the Norwegian Directorate of Public Construction and Property, as part of the Norwegian Ocean Technology Centre (NHTS) project which is currently under construction in Trondheim, Norway.
Csaba  P{\'a}kozdi at SINTEF Ocean has kindly supplied wavemaker benchmark data and the author is  grateful to Kontorbamse for continued academic support.

\bibliographystyle{abbrvnat} % abbrvnat,plainnat,unsrtnat
\bibliography{sintef_bib.bib}

\appendix 

\section{Projective mapping kernels} % stensils
\label{sec:mapChalikov}
At the core of present method is the ability to prescribe conformal maps that project surface or bathymetry states onto the domain interior.
These are all represented by the two transformation kernels % into a plane variable $z=x+\iiy$:
\begin{subequations}
	\begin{align}
		\CC{h}{\fun}(z) &
		= \sum_{j=-\MM}^\MM \FF_{\!j}(\fun) \frac{\ee^{\ii \kk_j(z+\ii h)}}{\cosh(\kk_jh)} 
		 =  \FF_{\!i}^{-1} \sbr{ \FF_{\!j}(\fun) \frac{2\, \ee^{- \kk_j y}}{1+\ee^{2\kk_jh}}},
		\label{eq:mapC}\\
		\SS{h}{\fun}(z) &
		= \sum_{j=-\MM}^\MM \FF_{\!j}(\fun) \frac{\ee^{\ii \kk_j(z+\ii h)}}{\delta_j-\sinh(\kk_jh)} 
		=  \FF_{\!i}^{-1} \sbr{\FF_{\!j}(\fun) \frac{2\,\ee^{-\kk_j y}}{1-\ee^{2\kk_jh}+2\delta_j}},
	\end{align}%
	\label{eq:mapSC}%
\end{subequations}%
$\FF\!_j$ being the discrete Fourier transform operator 
$  \FF_{\!j}\,(\fun) = \frac{1}{\N} \sum_{i=1}^\N \fun(x_i,t) \ee^{-\ii k_j x_i}$ and $  \FF_{\!i}^{-1}(\{\hat\fun_j\}) =  \sum_j\hat \fun_j \ee^{\ii k_j x_i}$ its inverse. 
Here,  $\delta_j$ deals with the zero-mode and equals one if $j=0$ and zero otherwise. 
The wavenumbers are defined as  $\kk_j=2\pi j/L$, $L$ being the horizontal domain length.
For simplicity, an odd number of points $ \N=2M+1$, uniformly spaced at a fixed depth, is here assumed.
If the number of points is even, then the kernels for the highest wavenumber $k=- \N\pi/L$  are
$\cosh\kk(h-\ii z)\sech\kk h$ and $\sinh \kk(h-\ii z)\csch\kk h$, respectively.

Assuming $\fun(x)$ is real, the kernels \eqref{eq:mapSC} %$\mapC_h$ and $\mapS_h$ 
have the following properties:
\begin{subequations}
	\begin{align}
		\CC{h}{\fun}(x) &= \fun(x)  -2\,\ii \sum_{j=1}^\MM \Im\Big[\FF_{\!j}(\fun) \ee^{\ii \kk_jx}\Big]\tanh\kk_j h,\label{eq:LProps:C0}\\
		\SS{h}{\fun}(x) &= \fun(x)  -2\,\ii \sum_{j=1}^\MM \Im\Big[\FF_{\!j}(\fun) \ee^{\ii \kk_jx}\Big]\coth\kk_jh;\label{eq:LProps:S0}\\
		\CC{h}{\fun}(x-\ii h) &= \mean{\fun} +2\,    \sum_{j=1}^\MM \Re\Big[\FF_{\!j}(\fun) \ee^{\ii \kk_jx}\Big]\sech\kk_jh,\label{eq:LProps:CH}\\
		\SS{h}{\fun}(x-\ii h) &=\mean{\fun}   -2\,\ii \sum_{j=1}^\MM \Im\Big[\FF_{\!j}(\fun) \ee^{\ii \kk_jx}\Big]\csch\kk_jh;\label{eq:LProps:SH}\\
		\lim_{h\to\infty}\CC{h}{\fun}(z) &=\lim_{h\to\infty}\SS{h}{\fun}(z)= \mean{\fun}  + 2 \sum_{j=-\MM}^{-1}\FF_{\!j}(\fun) \ee^{\ii \kk_j z}\label{eq:LProps:lim}
	\end{align}%
	\label{eq:LProps}%
\end{subequations}%
$\mean{\fun} = \FF_{\!0}(\fun)$ being the spatial mean.
This means that the real part of both kernels equals the input function itself at the line $y=0$, while $\mapC$ yields pure real values and $\mapS$ pure imaginary values at the line $y = -h$, except for a constant.
The equivalent projections that map $\mu$ to the bed reference line $y=-h$ are obtained with 
\[\{\CC{h}{\fun}(z^*-\ii h)\}^* = \CC{-h}{\fun}(z+\ii h)\]
and similar for $\mapS_h$.

An illustration of how the transformations can be implemented in practice is given in \autoref{list:CS} .
This also includes demonstration of derivation and domain mirroring.

\renewcommand{\^}{\slashHat}

\begin{lstlisting}[basicstyle=\ttfamily\small,label=list:CS,caption={Example function in MATLAB syntax returning projection mapping \eqref{eq:mapSC} (convolution operators) with option for derivation.}]
function convMu = convKernel(type,h,mu,z,n,MIRROR)
		nx0 = size(nu,1); x  = real(z); y = imag(z); dx = x(2)-x(1);
		if MIRROR
				nx2 = 2*(nx-1);
				y = y([1:nx,nx-1:-1:2],:);   % eq 8
				mu = mu([1:nx,nx-1:-1:2],:); % eq 8
		else
				nx2 = nx;
		end
		dk = 2*pi/(nx2*dx);
		k  = [0:ceil(nx2/2)-1,-floor(nx2/2):-1]'*dk;
		if type == 'C'
				sign = +1;
		elseif type == 'S'
				sign = -1;
		end
		kernel = 2./(exp(k.*y)+sign*exp(k.*(2*h+y))); kernel(1) = 1; % eq A.1
		if mod(nx2,2) == 0 % exception for largest even-number mode
				m = nx/2+1; y = y(m,:); km = k(m);
				kernel(m,:) = (exp(km.*(y+2*h))+sign.*exp(-km.*y))./(exp(2*km)*h)+sign);
		end
		convMu = ifft(fft(mu).*(1i.*k).^n.*kernel);% evaluation including derivative
		convMu = convMu(1:nx,:); % remove mirror plane 
end
\end{lstlisting}
\renewcommand{\^}[1]{^\mr{#1}}

\end{document}